\documentclass[journal]{IEEEtran}
\usepackage{graphicx}
\usepackage{cite}
\usepackage{picinpar}
\usepackage{amsmath}
\usepackage{stfloats}
\usepackage{url}
\usepackage{flushend}
\usepackage[latin1]{inputenc}
\usepackage{colortbl}
\usepackage{soul}
\usepackage{multirow}
\usepackage{pifont}
\usepackage{color}
\usepackage{alltt}
\usepackage[hidelinks]{hyperref}
\usepackage{enumerate}
\usepackage{siunitx}
\usepackage{breakurl}
\usepackage{epstopdf}
\usepackage{pbox}
\usepackage[caption=false]{subfig} 

\begin{document}
\title{Small-Signal Analysis of the Microgrid Secondary Control Considering a Communication Time Delay}

 \author{
 	\vskip 1em
         Ernane A. Alves Coelho,\emph{Member}, \emph{IEEE},
         Dan Wu,
         Josep M. Guerrero, \emph{Fellow Member}, \emph{IEEE},
         \\Juan C. Vasquez, \emph{Senior Member}, \emph{IEEE},
         Tomislav Dragi\v cevi\' c, \emph{Member}, \emph{IEEE},
         \\ \v Cedomir Stefanovi\' c , \emph{Member}, \emph{IEEE},
         Petar Popovski, \emph{Fellow Member}, \emph{IEEE}

	\thanks{
		
		Manuscript received September 20, 2015; revised December 25, 2015, February 25, 2016 and March 25, 2016; accepted April 11, 2016.
		The authors gratefully acknowledge the CAPES Foundation, Ministry of Education of Brazil,  for the financial support, grant number BEX9233/13-0. The work of C. Stefanovic was supported by the Danish Council for Independent Research, grant no. DFF-4005-00281. 
		
		D. Wu, J. M. Guerrero, J. C. Vasquez and T. Dragicevic are with the Department of Energy Technology, Aalborg University, Aalborg East DK-9220, Denmark (email: joz@et.aau.dk, juq@et.aau.dk).
		
		C.Stefanovic and P. Popovski are with the Department of Electronic Systems, Aalborg University, Aalborg East DK-9220, Denmark (email: cs@es.aau.dk, petarp@es.aau.dk).
		
		E. A. A. Coelho is with Universidade Federal de Uberlândia, Av. João Naves de Ávila 2121, Uberlândia, 38400-902, Brazil (email: ernane@ufu.br).		

	}
}

\maketitle
	
\begin{abstract}
This paper presents a small-signal analysis of an islanded microgrid composed of two or more voltage source inverters connected in parallel. The primary control of each inverter is integrated through internal current and voltage loops using PR compensators, a virtual impedance, and an external power controller based on frequency and voltage droops. The frequency restoration function is implemented at the secondary control level, which executes a consensus algorithm that consists of a load-frequency control and a single time delay communication network. The consensus network consists of a time-invariant directed graph and the output power of each inverter is the information shared among the units, which is affected by the time delay. The proposed small-signal model is validated through simulation results and experimental results. A root locus  analysis is presented that shows the behavior of the system considering control parameters and time delay variation.
\end{abstract}


{}

\definecolor{limegreen}{rgb}{0.2, 0.8, 0.2}
\definecolor{forestgreen}{rgb}{0.13, 0.55, 0.13}
\definecolor{greenhtml}{rgb}{0.0, 0.5, 0.0}

\section{Introduction}
\label{intro}

\IEEEPARstart{T}{he} growth in the applicability of the microgrid systems is a recent phenomenon, which consist of distributed systems where the sources and the loads are placed locally \cite{Guerrero2007}. The hierarchical control of a microgrid can be organized in three levels, primary, secondary and tertiary control as presented in \cite{Guerrero2011}. The primary control level, based on the droop control method, provides the power sharing between units,  but it applies the voltage and frequency deviations according to the load demand. Then, the functions of the voltage regulation and the frequency restoration, which need communication to operate, must be implemented at a secondary control level \cite{Guerrero2011, Bidram2013, Simpson2015}. The tertiary control manages the power flow  between the microgrid and the grid, considering the grid-connected operation. 

Several strategies for frequency and voltage restoration applied to the microgrid systems have been proposed \cite{Simpson2015, Fanghong2015, Ahumada2015}. In order to increase the system reliability by the addition of redundancy, the decentralized controller is preferred over the centralized one \cite{Simpson2015}. The secondary control can incorporates yet the cooperative characteristic, where each distributed source acts as an agent, which operates together with other agents to achieve a common goal. 

In \cite{Qobad2014} one notes that the distributed secondary control presents an improved performance, considering the communication latency, when compared to the central secondary control. The impact of communication delays on the secondary frequency control in an islanded microgrid is shown in \cite{Shichao2015}. However, in this case the frequency restoration is implemented in a centralized controller using a proportional integral compensator. In \cite{Ahumada2015}, robust control strategies for frequency restoration are implemented considering a variable and unknown time delay in data communication, but in this approach, a centralized secondary control is used, and additionally, the system control incorporates a PLL (Phase Locked Loop) to obtain the frequency at the bus loading.

The time delay effect on the system's stability has been the topic of investigation in several engineering applications by the use of delay differential equations (DDE). The spectrum analysis of DDE is more complicated than that of ordinary differential equations (ODE). The analytical solution is only possible in simple cases, where numerical approaches are used for practical systems \cite{Milano2012, Zhang2006}. In these cases, the effect of the time delays on power system stability is presented. 

This paper presents a small-signal modeling of a microgrid system operating in islanded mode, which presents a distributed control divided into primary and secondary levels. Frequency restoration based on a consensus algorithm is implemented in the secondary control level, which uses a specific control law and a data network. This data network presents a single time delay and its topology can be described using the graph theory. 

The contribution of this paper is in its presentation of an approach for building a DDE model for a microgrid with a single load bus, which allows for stability studies, taking into consideration the secondary and primary control parameters, the data network topology and the communication time delay. 

The rest of the paper is organized as follows. Section \ref{sec:1} presents the system control scheme. The proposed small-signal model of the system is presented in Section \ref{sec:2}. In order to validate the proposed model, simulation and experimental results are presented in Section \ref{sec_simula}. Section \ref{Ext_model} shows that it is a simple task to extend the model over to a system with more inverter units. Details about a communication system with constant time delay and the packet loss are presented in Section \ref{delay_packet}. Section \ref{conclusion} presents the conclusion of the present work.

\section{The Control Scheme}
\label{sec:1}

\begin{figure*}[!t]
\centering
\includegraphics[width=0.9\textwidth]{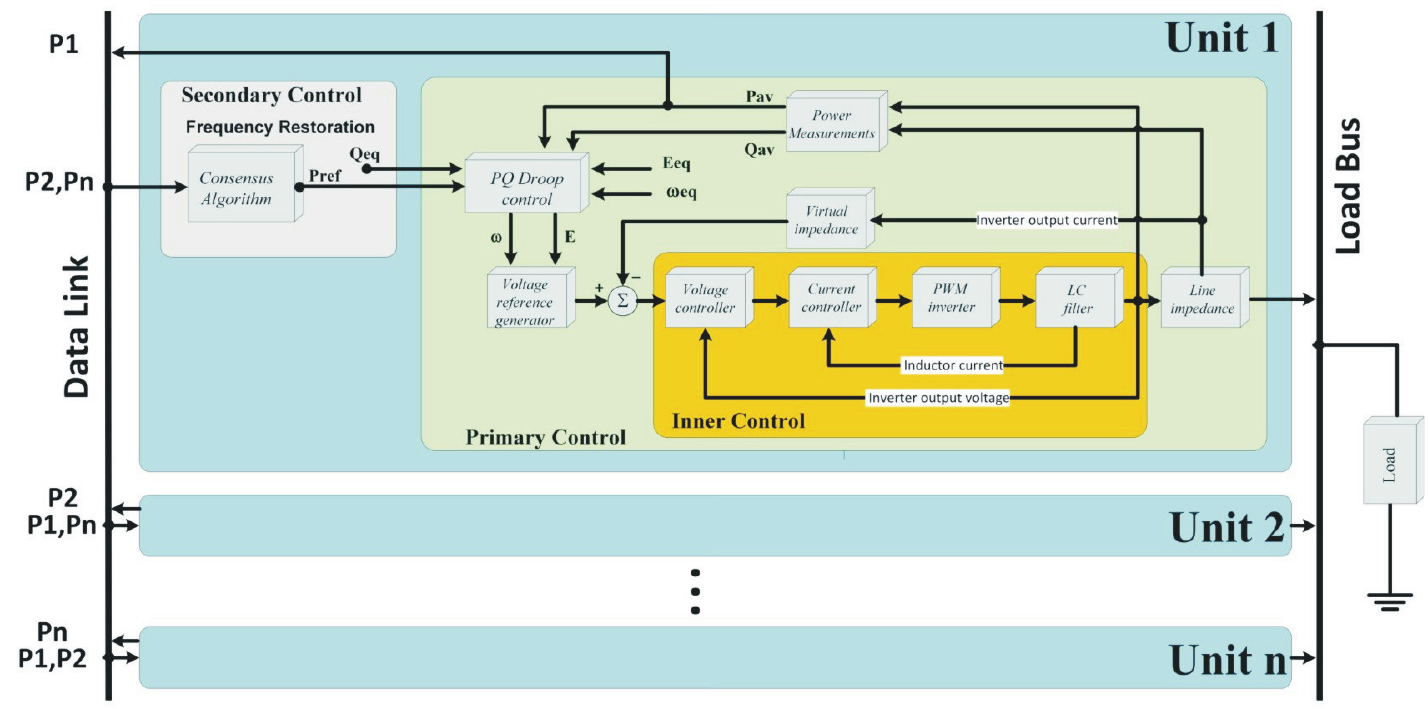}
\caption{The Microgrid Scheme}
\label{Sys_scheme}    
\end{figure*}

The complete scheme of the microgrid considered in this work is presented in Fig. \ref{Sys_scheme}. The microgrid is composed of an arbitrary number of inverter units. Each unit presents a hierarchical control, which integrates the inner control, the primary control and the secondary control \cite{Guerrero2011}. 

The inner control is composed of a current loop and an external voltage loop. In both loops  Proportional Resonant (PR) controllers are used  in $\alpha-\beta$ reference, considering the ideal function represented by  (\ref{ideal_PRes}), where $k_r$ is the proportional gain, $k_{res}$ is the resonant gain and $\omega$ is the frequency of the resonant pole, which in this case is equal to the grid frequency. To keep the resonant pole over the system frequency, the frequency reference provided by the primary control is used for frequency tracking. 

\begin{equation}
G_{PR}= k_r + k_{res}\frac{s}{s^2+\omega^2} 
\label{ideal_PRes}
\end{equation} 

In order to improve system stability, a virtual impedance is considered using the same implementation presented in \cite{Jinwei2011}, where the voltage drops over the virtual impedance $V_{v\alpha}$ and $V_{v\beta}$ are described by (\ref{Vvirtual}), being $Rv$ and $L_v$ the virtual resistance and inductance, respectively, and $I_{\alpha}$ and $I_{\beta}$ the inverter output currents in $\alpha-\beta$ reference.

\begin{equation}
\left[\begin{array}{c} V_{v\alpha}\\  V_{v\beta}\end{array} \right] = 
\left[\begin{array}{cc}     R_v        &   - \omega L_v \\
                     \omega L_v        &            R_v          \end{array} \right]
\left[\begin{array}{c} I_{\alpha}\\  I_{\beta}\end{array} \right]
\label{Vvirtual}
\end{equation} 
 
The primary control is based on the droop control method, which is capable of providing the active and reactive power sharing between the units without using communication, that is, only the local measurements are used. This control level is not capable of guaranteeing the equitable power sharing, since it is affected by the possible discrepancy of parameters between units, such as distinct line impedances. Besides that, the load affects the operational frequency and voltage.

The frequency and voltage droops used to control each inverter are described by (\ref{pxd}) and (\ref{qxv}),  respectively, these present the gains $k_p$ and $k_v $. $Q_{eq}$ is reactive power at the equilibrium point, where the inverter operates with the voltage amplitude $E_{eq}$ and a frequency $\omega_{eq}$. $P_{av}$ and $Q_{av}$ are the average active and reactive power measured by a data acquisition system in each inverter. $P_{ref}$ is the power reference of the frequency droop, a differentiated analysis to that presented in \cite{Coelho2002}, where $P_{ref}$ was a constant and equivalent to the active power $P_{eq}$ at the equilibrium point. Here it represents an input variable that will be defined by the secondary control. 
 
\begin{eqnarray}
\omega&=&\omega_{eq}-k_{p}(P_{av}- P_{ref})  \label{pxd}  \\
E&=&E_{eq}-k_{v}(Q_{av}- Q_{eq})            \label{qxv}   
\end{eqnarray}

The algorithms for active and reactive power measuring use a first order low-pass filter with cut-off frequency of $\omega_f$ , then the relationships between the instantaneous powers ($p$ and $q$) and average powers ($P_{av}$ and $Q_{av}$) measured by filters are:

\begin{eqnarray}
P_{av}&=&\frac{\omega_f}{s+\omega_f} p  \label{Pfilter}  \\
Q_{av}&=&\frac{\omega_f}{s+\omega_f} q  \label{Qfilter} 
\end{eqnarray}

In a microgrid system, the frequency restoration and the voltage regulation can be implemented by the secondary control, but a communication data link is necessary. In this work, a decentralized secondary control that performs only the frequency restoration function is presented. The control law implemented in each node of the distributed secondary control is described by (\ref{secondcntr_law}). The goal of this controller is to eliminate the difference between the power reference of the $i-th$ inverter to the active power supplied by the others, as presented in Section \ref{sec_2control}. The idea can be applied to an arbitrary number of units, but for the sake of simplicity, the model and its validation are presented considering only a three node system. Results for twelve-inverter system are presented in Section \ref{Ext_model}. The data link network that connects all units presents a single and constant time delay.

\begin{equation}
 P_{refi} = - k_{pri}\int \sum_{\substack{j=1 \\ j\neq i}}^{n} (P_{refi}-P_{avj})dt
\label{secondcntr_law}
\end{equation} 

\section{Small-Signal Analysis}
\label{sec:2}
In order to facilitate one's comprehension of the proposed small-signal model, the math development is divided into 5 sections. Initially, the small-signal analysis for the primary control in each inverter is presented. Considering the admittance nodal equation, the connection between the nodes provided by the power network and the loads is analyzed. The consensus algorithm for the secondary control and the data network are presented, on which all links present the same arbitrary time delay. Finally, the complete model is presented. The respective time delay is considered constant in this work. This assumption corresponds to practical real-time digital communication setups, in which inter-processing times of the received packets are made constant, by means of buffering and use of sequence-numbers/time-stamps contained in the packets \cite{RTP}. In other words, the delay between two packet arrivals, that inevitably varies, can be assumed to be made constant, and the delay assumed in the paper is the upper bound of the total allowed delay in the system, made equal for all communication links. Another communication impairment that arises in practice are the packet losses, which can also account for the cases when the packet delay exceeds the upper bound.  This impairment is not included in the analysis provided in this section. However, in the simulation results presented in Section \ref{delay_packet}, one also covers this aspect and shows that for a realistic packet loss probability that can be expected in practice, there is no significant difference between the results obtained by simulation and  the results obtained by the small-signal model in which the packet loss probability is not included. More details are presented in Section \ref{delay_packet}.

\subsection{Small-Signal Model for each Inverter Under The Primary Control}
\label{subsec:21}

Considering the linearization around the equilibrium point specified by $\omega_{eq}$, $E_{eq}$, $P_{eq}$, $Q_{eq}$ and the measuring filters described by (\ref{Pfilter}) and (\ref{Qfilter}) , one can rewrite (\ref{pxd}) and (\ref{qxv}) as:

\begin{eqnarray}
s\Delta\omega&=&-\omega_f \Delta\omega -k_p\omega_f \Delta p + k_p s\Delta P_{ref} \nonumber \\
             & & + k_p\omega_f \Delta P_{ref}  \label{dpxw}  \\
s\Delta E&=&-\omega_f \Delta E -k_v\omega_f\Delta q                \label{dqxv}  
\end{eqnarray}

It is important to keep in mind that $P_{ref}$ is a variable, therefore two extra terms are implied in (\ref{dpxw}) related to the deviation $\Delta P_{ref}$ and its derivative.

The analytical calculations for the inverter voltage are the same as those presented in \cite{Coelho2002}, those being, the inverter voltage $\vec{E}$, this can be written using a coordinate system with direct-axis and quadrature-axis:

\begin{equation}
\vec{E} = e_d+je_q = E cos(\delta) + j E sin(\delta)
\label{vetor_e}
\end{equation}
where:
\begin{equation}
\delta = arc tan(\frac{e_q}{e_d}) \label{delta1}  
\end{equation}

It is important to emphasize that $\delta$ is not the relative phase between output voltages of inverters connected to the system, but it is the absolute inverter voltage phase. Therefore, as one notes in Section \ref{sec_simula}, a redundant state and an eigenvalue at the origin \cite{Coelho2002} are implied. Besides this, as developed in \cite{Coelho2002}, the reference voltage of each inverter obtained by the voltage droop is considered as being equal to the inverter output voltage, that is, the inverters are considered as ideal voltage sources.

Linearizing (\ref{delta1}) for a given $e_d$ and $e_q$ defined by the equilibrium point:
\begin{equation}
\Delta\delta = \frac{\partial \delta}{\partial e_d} \Delta e_d + \frac{\partial \delta}{\partial e_q} \Delta e_q = m_d \Delta e_d + m_q \Delta e_q,
\label{delta2}
\end{equation}

where:
\begin{equation}
m_d=-\frac{e_q}{e_d^2+e_q^2} ,\quad m_q=\frac{e_d}{e_d^2+e_q^2} \label{mdmq}.  
\end{equation}

Since $\Delta\omega(s)= s \Delta\delta(s)$, then:
\begin{equation}
\Delta\omega= m_d \Delta \dot{e_d} + m_q \Delta \dot{e_q}.
\label{omega1}
\end{equation}

Considering that (\ref{mod_e}) defines the amplitude of the inverter voltage, its respective linearization around the equilibrium point can be obtained by equation (\ref{mod_e1}): 
\begin{equation}
E=|\vec{E}|=\sqrt{e_d^2 + e_q^2}
\label{mod_e}
\end{equation}

\begin{equation}
\Delta E = n_d \Delta e_d + n_q \Delta e_q
\label{mod_e1}
\end{equation}
where:
\begin{equation}
n_d=\frac{e_d}{\sqrt{e_d^2+e_q^2}} ,\quad n_q=\frac{e_q}{\sqrt{e_d^2+e_q^2}}.  \label{def_ndnq}
\end{equation}

Which implies that:
\begin{equation}
s\Delta E = n_d s\Delta e_d + n_q s\Delta e_q.
\label{mod_e2}
\end{equation}

Solving the equation system formed by (\ref{dqxv}), (\ref{omega1}), (\ref{mod_e1}) and (\ref{mod_e2}), isolating the derivatives $s\Delta e_d$ and $s\Delta e_q$, and considering (\ref{dpxw}), one obtains the state equation (\ref{state1}), which describes the behavior of the states $\Delta\omega$, $\Delta e_d$, and $\Delta e_q$ of the \emph{i-th} inverter in the neighborhood of the equilibrium point. As one can see, the input of the state equation includes a term which depends on the deviation of apparent power that the inverter is supplying, and the all other terms are related to the reference average power deviation and its derivative.
\begin{eqnarray}\nonumber
\left[\begin{array}{c}\Delta\dot{\omega_i}\\ \Delta\dot{e_{di}}\\ \Delta\dot{e_{qi}}\end{array}  \right] &=\left[M_i\right] \left[\begin{array}{c}\Delta\omega_i \\ \Delta e_{di} \\ \Delta e_{qi} \end{array} \right] + \left[B_{si}\right] \left[\begin{array}{c} \Delta p_i \\ \Delta q_i \end{array} \right]  \\ 
& + \left[B_{ri}\right] \left[\begin{array}{c} \Delta P_{refi} \end{array} \right] + \left[B_{di}\right] \left[\begin{array}{c} \dot{\Delta P_{refi}} \end{array} \right]
\label{state1}
\end{eqnarray}

or representatively:
\begin{eqnarray}\nonumber
\left[ \dot{\Delta X_{si}}  \right] & =\left[M_i\right] \left[\Delta X_{si} \right] + \left[B_{si}\right] \left[ \Delta S_i \right] \\
& + \left[B_{ri}\right] \left[\Delta Pref_i \right] + \left[B_{di}\right] \left[\dot{\Delta Pref_i} \right] 
\label{simb_state1}
\end{eqnarray}
where:
\begin{equation}
\left[M_i \right] = \left[\begin{array}{ccc}-\omega_f   &   0   &   0   \\ 
\frac{n_q}{m_d n_q - m_q n_d}& \frac{m_q n_d \omega_f}{m_d n_q - m_q n_d}& \frac{m_q n_q \omega_f}{m_d n_q - m_q n_d} \\
\frac{n_d}{m_q n_d - m_d n_q}& \frac{m_d n_d \omega_f}{m_q n_d - m_d n_q}& \frac{m_d n_q \omega_f}{m_q n_d - m_d n_q}  \end{array} \right]
\label{def_mi}
\end{equation}

\begin{equation}
\left[B_{si} \right] = \left[\begin{array}{cc} -k_p\omega_f   &   0  \\ 
   0  &  \frac{k_v m_q\omega_f}{m_d n_q - m_q n_d}  \\
   0  &  \frac{k_v m_d\omega_f}{m_q n_d - m_d n_q} \end{array} \right]
\label{def_bsi}
\end{equation}

\begin{equation}
\left[B_{ri} \right] = \left[\begin{array}{c}  k_p\omega_f    \\ 
                                                0             \\
                                                0      \end{array} \right]
\label{def_bri}
\end{equation}

\begin{equation}
\left[B_{di} \right] = \left[\begin{array}{c}  k_p   \\ 
                                                0    \\
                                                0      \end{array} \right]
\label{def_bdi}
\end{equation}

\subsection{Small-Signal Model for The Entire Microgrid Under The Primary Control}
\label{subsec:22}
\begin{figure}[!t]
\centering
\includegraphics[width=0.65\columnwidth]{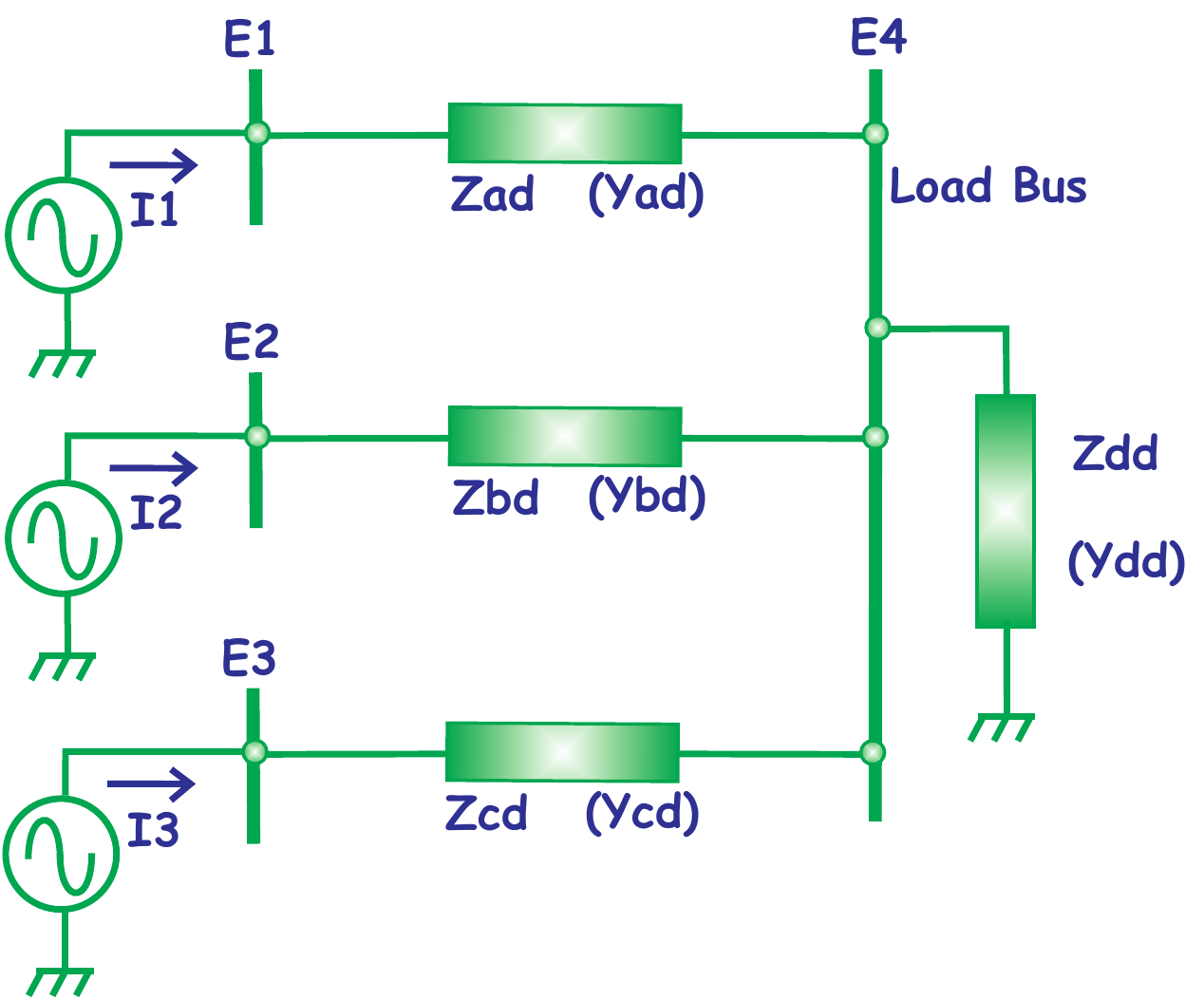}
\caption{Parallel-connected inverters in an islanded microgrid}
\label{Sys_3invs}    
\end{figure}

The principle used to develop the model can be applied to a microgrid with an arbitrary number of nodes. However, in order to facilitate this development, an islanded microgrid will be examined; this is composed of three inverters connected in parallel to a common load bus, as visualized in Fig. \ref{Sys_3invs}. 

In order to simplify the power network analysis, the effect of frequency variation over the frequency-dependent loads will be neglected, that is, the network reactances will be considered constant. This assumption can be considered reasonable because the droop controllers are designed to apply low deviations along the system frequency. It is important to keep in mind that the higher the system frequency range, the lower the precision of this modeling will be.

\begin{figure}[!t]
\centering
\includegraphics[width=0.75\columnwidth]{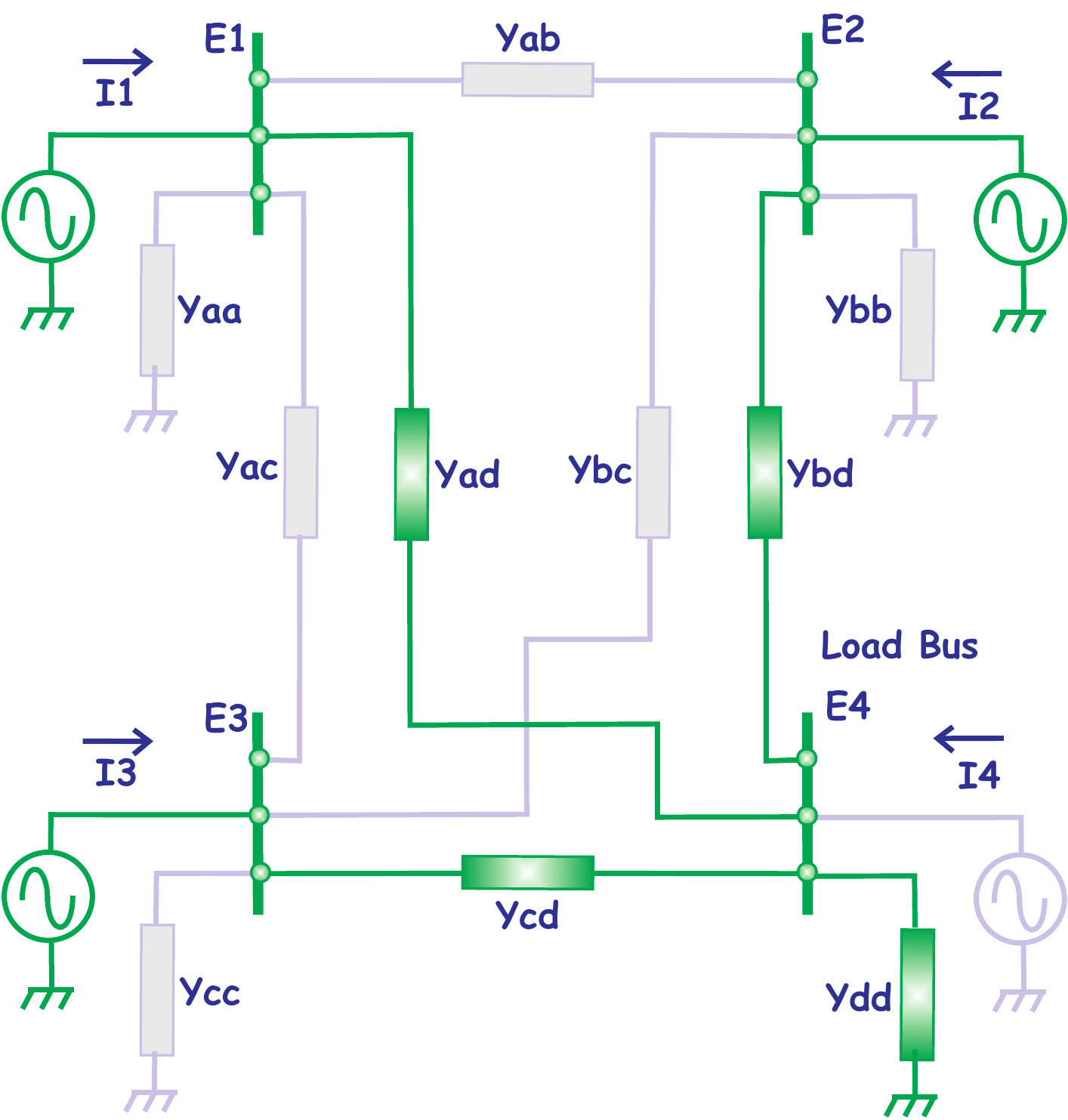}
\caption{Relation Between the Common Load Bus and Regular Networked Microgrids}
\label{Y4bars}    
\end{figure}

Therefore, neglecting the frequency variations, the nodal admittance equation for the islanded microgrid presented in Fig. \ref{Sys_3invs} can be obtained considering the regular networked microgrid shown in Fig. \ref{Y4bars}, where the gray admittances are null and there is no inverter connected in the load bus. 

Hence, the nodal equation of the islanded microgrid is (\ref{nodal_eq}), which in its representative form is (\ref{snodal_eq}). 

\scriptsize  
\begin{equation}
\left[\begin{array}{c}\vec{I_1} \\ \vec{I_2} \\ \vec{I_3} \\ \vec{I_4}\end{array}\right] =  
\left[\begin{array}{cccc} Y_{ad}  &     0   &    0    &   -Y_{ad}  \\ 
                            0     &  Y_{bd} &    0    &   -Y_{bd}  \\
                            0     &     0   &  Y_{cd} &   -Y_{cd}  \\
                         -Y_{da}  & -Y_{db} & -Y_{dc} &  Y_{da}+Y_{db}+Y_{dc}+Y_{dd} \end{array} \right] 
\left[\begin{array}{c}\vec{E_1} \\ \vec{E_2} \\ \vec{E_3} \\ \vec{E_4} \end{array}\right] 
\label{nodal_eq}
\end{equation}
\normalsize

\begin{equation}
\left[\begin{array}{c}\vec{I_{1234}} \end{array}\right] = 
\left[\begin{array}{c} Y      \end{array} \right] 
\left[\begin{array}{c}\vec{E_{1234}} \end{array}\right] \\.
\label{snodal_eq}
\end{equation}

Since there is no power injection on node 4 and the all power consumption is represented by the respective shunt load included in the admittance matrix $Y$, the voltage at node 4 is a linear combination of the voltage on the other 3 nodes. Thus, we can eliminate node 4 by considering (\ref{Y4to3}), which is derived from (\ref{nodal_eq}), considering $\vec{I_4}=0$. 

\begin{equation}
\left[\begin{array}{c}\vec{E_1} \\ \vec{E_2} \\ \vec{E_3} \\ \vec{E_4}\end{array}\right] = 
\left[\begin{array}{ccc}   1         &     0        &    0      \\ 
                           0         &     1        &    0      \\
                           0         &     0        &    1      \\
                         Y_{da}/Y_t  &  Y_{db}/Y_t  &  Y_{dc}/Y_t \end{array} \right] 
\left[\begin{array}{c}\vec{E_1} \\ \vec{E_2} \\ \vec{E_3} \end{array}\right] \\
\label{Y4to3}
\end{equation}

or representatively: 
\begin{equation}
\left[\begin{array}{c}\vec{E_{1234}} \end{array}\right] = 
\left[\begin{array}{c} T_{4to3}      \end{array} \right] 
\left[\begin{array}{c}\vec{E_{123}} \end{array}\right] \\,
\label{sY4to3}
\end{equation}

where $Y_t=Y_{da}+Y_{db}+Y_{dc}+Y_{dd}$.

Then, the admittance nodal equation of the three-inverter system shown in Fig. \ref{Sys_3invs} is: 

\begin{equation}
\left[\begin{array}{c}\vec{I_1} \\ \vec{I_2} \\ \vec{I_3} \end{array}\right] = 
\left[\begin{array}{c} Y_s \end{array} \right] 
\left[\begin{array}{c}\vec{E_1} \\ \vec{E_2} \\ \vec{E_3}  \end{array}\right] \\
\label{nodal_eq3inv}
\end{equation}

where the matrix $[Y_s]$ is the submatrix(1:3,1:3) of the product $[Y][ T_{4to3}]$.

Converting the complex equation (\ref{nodal_eq3inv}) to its real form:

\scriptsize 
\begin{eqnarray}
\left[\begin{array}{c} i_{d1}\\  i_{q1}\\  i_{d2}\\  i_{q2} \\ i_{d3}\\  i_{q3}\end{array}\right] = 
\left[\begin{array}{cccccc} G_{11} & -B_{11} & G_{12} & -B_{12} & G_{13} & -B_{13}  \\ 
                            B_{11} &  G_{11} & B_{12} &  G_{12} & B_{13} &  G_{13}  \\
                            G_{21} & -B_{21} & G_{22} & -B_{22} & G_{23} & -B_{23}  \\ 
                            B_{21} &  G_{21} & B_{22} &  G_{22} & B_{23} &  G_{23}  \\
                            G_{31} & -B_{31} & G_{32} & -B_{32} & G_{33} & -B_{33}  \\ 
                            B_{31} &  G_{31} & B_{32} &  G_{32} & B_{33} &  G_{33}  \\  \end{array} \right] 
\left[\begin{array}{c} e_{d1}\\  e_{q1}\\  e_{d2}\\  e_{q2}\\  e_{d3}\\  e_{q3} \end{array}\right]  
\label{nodal_real}
\end{eqnarray}
\normalsize
where:
\begin{equation}
  Y_{sij}=G_{ij}+jB_{ij}.                  
\label{admittance}
\end{equation}

Linearizing  (\ref{nodal_real}), one obtains (\ref{sim_nodal2}).

\begin{equation}
\left[ \Delta i \right] = \left[ Y_s \right]  \left[  \Delta e \right]
\label{sim_nodal2}  
\end{equation}

Considering the expressions used for calculating the active and reactive power for the \emph{i-th} inverter using a \emph{d-q} orthogonal coordinate system, one has:

\begin{eqnarray}
p_i&=&e_{di} i_{di} + e_{qi} i_{qi}  \label{p_i}  \\
q_i&=&e_{di} i_{qi} - e_{qi} i_{di}  \label{q_i}  
\end{eqnarray}

Considering the system presented in Fig. \ref{Sys_3invs} and linearizing (\ref{p_i}) and (\ref{q_i}) , one obtains (\ref{deltaPQ}), which describes the deviations of the active and reactive power around the equilibrium point.

\scriptsize
\begin{eqnarray} \nonumber 
\left[\begin{array}{c}\Delta p_1 \\ \Delta q_1 \\ \Delta p_2 \\ \Delta q_2\\  \Delta p_3 \\ \Delta q_3\end{array}\right] = 
\left[\begin{array}{cccccc} i_{d1} &  i_{q1}  &    0    &    0    &    0    &    0    \\ 
                            i_{q1} & -i_{d1}  &    0    &    0    &    0    &    0    \\
                               0   &     0    &  i_{d2} &  i_{q2} &    0    &    0    \\ 
                               0   &     0    &  i_{q2} & -i_{d2} &    0    &    0    \\
                               0   &     0    &    0    &     0   &  i_{d3} &  i_{q3} \\ 
                               0   &     0    &    0    &     0   &  i_{q3} & -i_{d3}   \end{array} \right] 
\left[\begin{array}{c}\Delta e_{d1}\\ \Delta e_{q1}\\ \Delta e_{d2}\\ \Delta e_{q2} \\ \Delta e_{d3}\\ \Delta e_{q3}\end{array}\right] \nonumber \\ + 
\left[\begin{array}{cccccc} e_{d1} &  e_{q1}  &     0    &    0    &     0    &    0    \\ 
                           -e_{q1} &  e_{d1}  &     0    &    0    &     0    &    0    \\
                               0   &     0    &   e_{d2} &  e_{q2} &     0    &    0    \\ 
                               0   &     0    &  -e_{q2} &  e_{d2} &     0    &    0    \\
                               0   &     0    &     0    &    0    &  e_{d3}  &  e_{q3} \\ 
                               0   &     0    &     0    &    0    & -e_{q3}  &  e_{d3}   \end{array} \right] 
\left[\begin{array}{c}\Delta i_{d1}\\ \Delta i_{q1}\\ \Delta i_{d2}\\ \Delta i_{q2} \\ \Delta i_{d3}\\ \Delta i_{q3}\end{array}\right]                          
\label{deltaPQ}
\end{eqnarray}
\normalsize

Equation (\ref{deltaPQ}) can be written representatively as:
\begin{equation}
[\Delta S]=[I_s][\Delta e]+[E_s][\Delta i].
\label{sim_deltaPQ}
\end{equation}

Substituting (\ref{sim_nodal2}) into (\ref{sim_deltaPQ}), then:
\begin{equation}
[\Delta S]=\left([I_s]+[E_s][Y_s]\right) [\Delta e].
\label{delta_S}
\end{equation}

The state equation that represents the system shown in Fig. \ref{Sys_3invs} can be derived from (\ref{state1}), this represents each inverter separately. Thus resulting in the state equation (\ref{fullstate_eq}).

\begin{eqnarray}\nonumber 
\left[\begin{array}{c}\dot{\Delta\omega_1}\\ \dot{\Delta e_{d1}}\\ \dot{\Delta e_{q1}} \\ 
                      \dot{\Delta\omega_2}\\ \dot{\Delta e_{d2}}\\ \dot{\Delta e_{q2}} \\
                      \dot{\Delta\omega_3}\\ \dot{\Delta e_{d3}}\\ \dot{\Delta e_{q3}} \end{array} \right] &=& 
\left[\begin{array}{ccc} M_1 &   &  \\   & M_2 &   \\   &   & M_3 \end{array}\right] 
\left[\begin{array}{c} \Delta\omega_1\\ \Delta e_{d1}\\ \Delta e_{q1} \\ 
                       \Delta\omega_2\\ \Delta e_{d2}\\ \Delta e_{q2} \\ 
                       \Delta\omega_3\\ \Delta e_{d3}\\ \Delta e_{q3} \end{array} \right] \\ 
\nonumber 
&+&
\left[\begin{array}{ccc} B_{s1} &         &        \\ 
                               &  B_{s2} &        \\
                               &         & B_{s3}    \end{array}\right] 
\left[\begin{array}{c} \Delta p_1 \\ \Delta q_1 \\ 
                       \Delta p_2 \\ \Delta q_2 \\
                       \Delta p_3 \\ \Delta q_3\end{array}\right]\\
\nonumber                        
&+& 
\left[\begin{array}{ccc} B_{r1} &         &        \\ 
                               &  B_{r2} &        \\ 
                               &         &  B_{r3}\end{array}\right] 
\left[\begin{array}{c} \Delta P_{ref1} \\ \Delta P_{ref2} \\ \Delta P_{ref3}\end{array}\right] \\
&+& 
\left[\begin{array}{ccc} B_{d1} &         &     \\ 
                                &  B_{d2} &     \\
                                &         & B_{d3} \end{array}\right] 
\left[\begin{array}{c}\dot{\Delta P_{ref1}} \\ \dot{\Delta P_{ref2}} \\ \dot{\Delta P_{ref3}}\end{array}\right]
\label{fullstate_eq}
\end{eqnarray}

Or representatively as:
\begin{eqnarray}\nonumber 
[\Delta\dot{X_s}]&=&[M_s][\Delta X_s]+[B_{ss}][\Delta S] \\
& & +[B_{rs}][\Delta P_{refs}]+[B_{ds}][\dot{\Delta P_{refs}}]
\label{simfullstate_eq}
\end{eqnarray}

Then, combining (\ref{delta_S}) and (\ref{simfullstate_eq}):
\begin{eqnarray}\nonumber 
[\Delta \dot{X_s}]&=&[M_s][\Delta X_s]+ [B_{ss}] \left([I_s]+[E_s][Y_s]\right) [\Delta e] \\ 
 & & +[B_{rs}][\Delta P_{refs}]+[B_{ds}][\dot{\Delta P_{refs}}] 
\label{def_Xs} 
\end{eqnarray}

One observes that the relation between $\Delta e$  and the state vector $\Delta X_s$ is:

\scriptsize 
\begin{eqnarray}
\left[\begin{array}{c} \Delta e_{d1}\\ \Delta e_{q1} \\ 
                       \Delta e_{d2}\\ \Delta e_{q2} \\ 
                       \Delta e_{d3}\\ \Delta e_{q3} \end{array} \right] = 
\left[\begin{array}{ccccccccc}0&1&0&0&0&0&0&0&0\\
                              0&0&1&0&0&0&0&0&0\\
                              0&0&0&0&1&0&0&0&0\\
                              0&0&0&0&0&1&0&0&0\\
                              0&0&0&0&0&0&0&1&0\\
                              0&0&0&0&0&0&0&0&1\end{array} \right] 
\left[\begin{array}{c} \Delta\omega_1\\ \Delta e_{d1}\\ \Delta e_{q1} \\ 
                       \Delta\omega_2\\ \Delta e_{d2}\\ \Delta e_{q2} \\ 
                       \Delta\omega_3\\ \Delta e_{d3}\\ \Delta e_{q3} \end{array} \right]
\label{const_ke}
\end{eqnarray}
\normalsize

Which representatively is:
\begin{equation}
[\Delta e]=[K_e][\Delta X_s] 
\label{simconst_ke}
\end{equation}

Substituting (\ref{simconst_ke}) for (\ref{def_Xs}), then:
\begin{eqnarray}\nonumber 
[\Delta \dot{X_s}]&=&[M_s][\Delta X_s]+[B_{ss}] \left([I_s]+[E_s][Y_s]\right) [K_e][\Delta X_s]\\ 
& & +[B_{rs}][\Delta P_{refs}]+[B_{ds}][\dot{\Delta P_{refs}}] 
\label{def_Xs2}
\end{eqnarray}

After some algebraic manipulations, we can obtain the state equation (\ref{stateeqsys}), which describes the behavior of the system considering a given initial condition in the neighborhood of the equilibrium point and the input deviations $\Delta P_{refs}$ and its derivatives. If the inputs of the state equation are considered null, the small-signal analysis falls into the particular case presented in \cite{Coelho2002}, where a secondary control level is not considered. 

\begin{eqnarray}\nonumber 
[\Delta \dot{X_s}]&=&\left([M_s]+[B_{ss}] \left( [I_s]+[E_s][Y_s] \right) [K_e]\right)[\Delta X_s]\\ 
& & +[B_{rs}][\Delta P_{refs}]+[B_{ds}][\dot{\Delta P_{refs}}] 
\label{stateeqsys}
\end{eqnarray}

\subsection{Small-Signal Model for The Entire Microgrid Under The Secondary Control}
\label{sec_2control}
The goal of the secondary control in this work is to keep the system frequency over the nominal value in spite of the load variation, but concomitantly keeping the equitable active power sharing, that is, its function is the frequency restoration. Thus, in order to perform this function, the secondary control modifies the power reference $P_{refi}$ of the frequency droop in each inverter. 

\begin{figure}[!t]
\centering
\includegraphics[width=0.5\columnwidth]{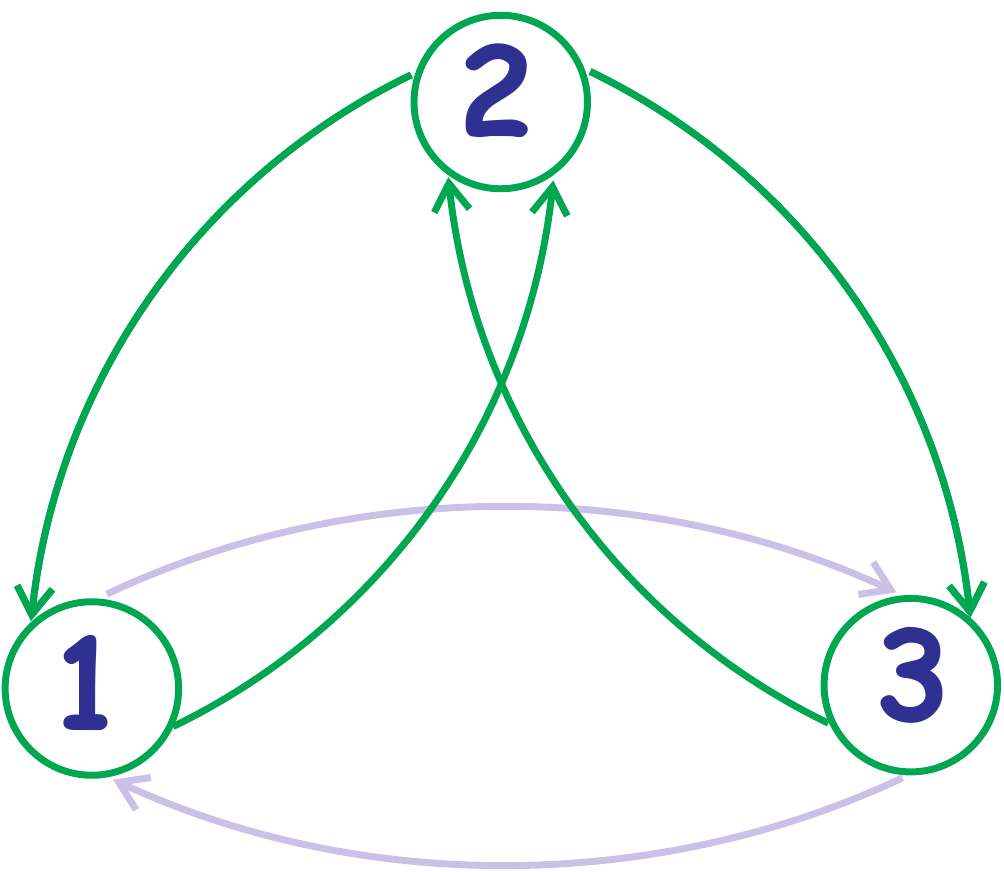}
\caption{Directed Graph for Secondary Control}
\label{Consensus_Sys}    
\end{figure}

The islanded microgrid presented in Fig. \ref{Sys_3invs} can be considered as a power network where there is a consensus to provide the power sharing, and where the frequency and voltage droops are the distributed controllers. This consensus keeps the system stable and in steady state all inverters operate at the same frequency, not necessarily the nominal frequency. The load sharing and the equilibrium frequency depend on  the load and the setpoints of the reference power in each inverter. Thus, another network will be used for implementing the frequency restoration that being a data link network. This new network can be presented in several topologies. A strongly connected example \cite{Rezasaber2004} is shown in Fig. \ref{Consensus_Sys}, where only three inverters are considered.  

The data link network in Fig. \ref{Consensus_Sys} is a directed graph where the inverters are the vertices and the directional data links are the edges. In this work, a not strongly connected directed network will be considered, so the data links shown in gray will be neglected. Thus, the adjacency matrix $A_g$ and the degree matrix $D_g$ of the directed graph presented in Fig. \ref{Consensus_Sys} are:

\begin{equation}
\left[ A_g \right] = 
\left[\begin{array}{cccc}0&1&0\\
                         1&0&1\\
                         0&1&0   \end{array} \right] , \quad \left[ D_g \right] = 
\left[\begin{array}{cccc}1&0&0\\
                         0&2&0\\
                         0&0&1   \end{array} \right] 
\label{adjacency_degree_m}
\end{equation} 

It is possible to implement different types of consensus algorithm into the secondary control level. For example, to implement the active power sharing an average-consensus algorithm can be used. This kind of consensus can be represented by (\ref{consensus_eq}) \cite{Rezasaber2004, Rezasaber, Linyu}, where $x$ is the state vector of the system, $\mathbf{L}$ is the Laplacian matrix of the graph defined by (\ref{Laplacian_def}).

\begin{equation}
\dot{x}=-C(\mathbf{D_g}-\mathbf{A_g})x=-C\mathbf{L}x
\label{consensus_eq}
\end{equation}

\begin{equation}
\mathbf{L}=\mathbf{D_g}-\mathbf{A_g}.
\label{Laplacian_def}
\end{equation} 

In this case, the distributed control law can be represented by (\ref{outcntr_law}), considering an unweighted graph, where $C$ is a constant called the diffusion constant, which  affects the convergence rate \cite{Linyu}.

\begin{equation}
 P_{refi} = - C\int \sum_{\substack{j=1 \\ j\neq i}}^{n} (P_{avi}-P_{avj})dt
\label{outcntr_law}
\end{equation} 

The consensus algorithm implemented using the distributed controller represented by (\ref{outcntr_law}) is capable of keeping the equitable active power sharing in spite of load variation, but it does not guarantee the operation at the nominal frequency. Therefore, in order to meet both requirements, in this work the control law for the secondary control implemented in each inverter is described by (\ref{scntr_law}), this corresponds to the distributed controller implemented in the multi-agent system represented by the graph in Fig. \ref{Consensus_Sys}, where $k_{pri}$ is the integral gain of the controller in each inverter. 

\begin{equation}
 P_{refi} = - k_{pri}\int \sum_{\substack{j=1 \\ j\neq i}}^{n} (P_{refi}-P_{avj})dt
\label{scntr_law}
\end{equation} 

The terms to be included in the summation presented in (\ref{scntr_law}) depend on the topology of the data link network, that is, the existence of an outgoing edge from vertex $j$, which is incident on vertex $i$, implying the term $(P_{refi}-P_{avj})$ in the summation. It is assumed that all vertex has at least one incoming edge, which implies that all distributed controllers have at least one term in the summation. Then, considering the data link network as the graph described by the green line edges (Fig. \ref{Consensus_Sys}), the linearization of the control law shown in (\ref{scntr_law}) is:

\tiny
\begin{eqnarray}\nonumber 
\left[\begin{array}{c}\dot{\Delta P_{ref1}}\\  
                      \dot{\Delta P_{ref2}}\\ 
                      \dot{\Delta P_{ref3}} \end{array} \right] &=& - 
\left[\begin{array}{ccc} k_{pr1} &  0     &    0    \\  
                           0     & k_{pr2}&    0    \\ 
                           0     &  0     &  k_{pr3} \end{array} \right]                       
\left[\begin{array}{ccc}   1  &  0  &  0    \\  
                           0  &  2  &  0    \\ 
                           0  &  0  &  1 \end{array} \right] 
\left[\begin{array}{c} \Delta P_{ref1}\\  
                       \Delta P_{ref2}\\ 
                       \Delta P_{ref3} \end{array} \right]                          
 \\ 
 &+& \left[\begin{array}{ccc} k_{pr1} &   0    &    0    \\  
                                0     & k_{pr2}&    0    \\ 
                                0     &   0    &  k_{pr3} \end{array} \right] 
     \left[\begin{array}{ccc}   0  &  1  &  0    \\  
                                1  &  0  &  1    \\ 
                                0  &  1  &  0 \end{array} \right] 
     \left[\begin{array}{c} \Delta P_{av1}\\  
                            \Delta P_{av2}\\ 
                            \Delta P_{av3} \end{array} \right]
\label{delta_cntr_law} 
\end{eqnarray}
\normalsize

or in its representative form:

\begin{equation}
[ \dot{\Delta P_{refs}} ] = -[k_{prs}][D_g][ \Delta P_{refs} ]+[k_{prs}][A_g][ \Delta P_{avs}].
\label{sdelta_cntr_law}
\end{equation} 

If a distinct graph is considered with different edges from those highlighted in Fig. \ref{Consensus_Sys}, to obtain a new control law, it is  necessary only to change the degree matrix $D_g$ and adjacency matrix $A_g$ in (\ref{sdelta_cntr_law}). It is important to emphasize that no loop is considered in the network graph, that is, the term $P_{refi}-P_{avi}$ is not presented in the summation of (\ref{scntr_law}). This would be an option for keeping the nominal frequency in case of only one inverter or vertex remaining in operation, but in fact, no consensus is necessary if only one vertex is presented, since the nominal frequency could be imposed by the controller. Thus, as the loops are not considered in simple graphs, they will not be considered here either.

\subsection{Time Delay on The Secondary Control}
\label{sec_2control_delay}

Equation (\ref{sdelta_cntr_law}) represents the distributed controller in each inverter if no time delay is present in the data communication link. However, in this work, a constant time delay $t_d$ will be considered in each data communication link  represented by the edges on the network graph. Then, (\ref{sdelta_cntr_law_td}) must replace (\ref{sdelta_cntr_law}).

\begin{eqnarray}\nonumber 
[ \dot{\Delta P_{refs}}(t) ] &=& -[k_{prs}][D_g][ \Delta P_{refs}(t) ] \\
& & +[k_{prs}][A_g][ \Delta P_{avs}(t-t_d)].
\label{sdelta_cntr_law_td}
\end{eqnarray}

Substituting (\ref{sdelta_cntr_law_td}) in (\ref{stateeqsys}), it is possible to eliminate the input derivative term in the small-signal model for the islanded microgrid under the primary level control. Then, after some algebraic manipulations:

\begin{eqnarray}\nonumber 
[\dot{\Delta X_s}(t)]&=&\left([M_s]+[B_{ss}] \left( [I_s]+[E_s][Y_s] \right) [K_e]\right)[\Delta X_s(t)]\\
\nonumber 
& & + ([B_{rs}] - [B_{ds}][k_{prs}][D_g])[ \Delta P_{refs}(t) ] \\ 
& & + [B_{ds}][k_{prs}][A_g][ \Delta P_{avs}(t-t_d)]). 
\label{stateeq_xs}
\end{eqnarray}

It is important to keep in mind that the states in vector $\Delta X_s$ and vector $\Delta P_{refs}$ imply local feedbacks and no data communication link is necessary. Only the inverter output power measurement is sent from one vertex to the other using the data communication link, which is affected by the time delay $t_d$.

According to (\ref{Pfilter}) the relation between the deviations from average active power and instantaneous power in each inverter that integrates the network is:

\begin{eqnarray}\nonumber 
\left[\begin{array}{c}\dot{\Delta P_{av1}}\\  
                      \dot{\Delta P_{av2}}\\ 
                      \dot{\Delta P_{av3}} \end{array} \right] = &-&
\left[\begin{array}{ccc} \omega_{f1} &      0       &    0    \\  
                              0      & \omega_{f2}  &    0    \\ 
                              0      &      0       & \omega_{f3} \end{array} \right]      \left[\begin{array}{c} \Delta P_{av1}\\  
                       \Delta P_{av2}\\ 
                       \Delta P_{av3} \end{array} \right]                           
\\ 
 &+& \left[\begin{array}{ccc} \omega_{f1} &     0       &     0    \\  
                                  0       & \omega_{f2} &     0    \\ 
                                  0       &     0       & \omega_{f3} \end{array} \right]
     \left[\begin{array}{c} \Delta p_1\\  
                            \Delta p_2\\ 
                            \Delta p_3 \end{array} \right].
\label{pfilter3} 
\end{eqnarray}

Or representatively:

\begin{equation}
[\dot{\Delta P_{avs}}(t)] = -[\omega_{fs}] [\Delta P_{avs}(t)] + [\omega_{fs}] [\Delta p_s(t)]. 
\label{spfilter3}
\end{equation}

It is possible to represent the vector $\Delta p_s$ as a function of the vector the $\Delta S$, thus it follows that:

\begin{equation}
\left[\begin{array}{c} \Delta p_1\\  
                       \Delta p_2\\ 
                       \Delta p_3 \end{array} \right] = 
\left[\begin{array}{cccccc}1&0&0&0&0&0\\
                           0&0&1&0&0&0\\
                           0&0&0&0&1&0\end{array} \right] 
\left[\begin{array}{c}\Delta p_1 \\ \Delta q_1 \\ 
                      \Delta p_2 \\ \Delta q_2\\  
                      \Delta p_3 \\ \Delta q_3\end{array}\right],
\label{p_delta_s} 
\end{equation}

which in its representative form is:

\begin{equation}
[\Delta p_s] = [k_{ps}] [\Delta S]. 
\label{sp_delta_s}
\end{equation}

Applying (\ref{delta_S}), (\ref{simconst_ke}) and (\ref{sp_delta_s}) into (\ref{spfilter3}), we obtain (\ref{spav}).

\begin{eqnarray}\nonumber 
[\dot{\Delta P_{avs}}(t)] &=& -[\omega_{fs}] [\Delta P_{avs}(t)] + [\omega_{fs}][k_{ps}]([I_s] \\
& & + [E_s][Y_s])[K_e][\Delta X_s(t)]. 
\label{spav}
\end{eqnarray}

\subsection{Small-Signal Model for The Entire System - a DDE Model}
\label{sec_dde}

Considering (\ref{sdelta_cntr_law_td}), (\ref{stateeq_xs}) and (\ref{spav}), it is possible to write the state equation (\ref{sysstate_eq}) which corresponds to the small-signal model for the whole system, where the vectors $\Delta X_s$, $\Delta P_{avs}$ and  $\Delta P_{refs}$ are the components of the new state vector $\Delta X$. 


\begin{figure*}[!t]
\newcounter{MYtempeqncnt}
\setcounter{MYtempeqncnt}{\value{equation}}

\small
\begin{eqnarray}\nonumber 
\left[\begin{array}{c}
\dot{\Delta\omega_1}(t)\\ \dot{\Delta e_{d1}}(t)\\  \dot{\Delta e_{q1}}(t) \\ 
\dot{\Delta\omega_2}(t)\\ \dot{\Delta e_{d2}}(t)\\  \dot{\Delta e_{q2}}(t) \\
\dot{\Delta\omega_3}(t)\\ \dot{\Delta e_{d3}}(t)\\  \dot{\Delta e_{q3}}(t) \\
\dot{\Delta P_{av1}}(t)\\ \dot{\Delta P_{av2}}(t)\\ \dot{\Delta P_{av3}}(t) \\
\dot{\Delta P_{ref1}}(t)\\ \dot{\Delta P_{ref2}}(t)\\ \dot{\Delta P_{ref3}}(t) 
 \end{array} \right] = 
\left[\begin{array}{ccccccccccccccc}
  \multicolumn{9}{c|}{ } & 0 & 0 & 0 &\multicolumn{3}{|c}{ }\\ 
  \multicolumn{9}{c|}{ } & 0 & 0 & 0 &\multicolumn{3}{|c}{ }\\ 
  \multicolumn{9}{c|}{ } & 0 & 0 & 0 &\multicolumn{3}{|c}{ }\\ 
  \multicolumn{9}{c|}{M_s} & 0 & 0 & 0 &\multicolumn{3}{|c}{B_{rs}}\\ 
  \multicolumn{9}{c|}{+B_{ss}(I_s+E_s Y_s)k_e } & 0 & 0 & 0 &\multicolumn{3}{|c}{-B_{ds}k_{prs}D_g}\\ 
  \multicolumn{9}{c|}{ } & 0 & 0 & 0 &\multicolumn{3}{|c}{ }\\ 
  \multicolumn{9}{c|}{ } & 0 & 0 & 0 &\multicolumn{3}{|c}{ }\\ 
  \multicolumn{9}{c|}{ } & 0 & 0 & 0 &\multicolumn{3}{|c}{ }\\ 
  \multicolumn{9}{c|}{ } & 0 & 0 & 0 &\multicolumn{3}{|c}{ }\\  \hline 
  \multicolumn{9}{c|}{} & \multicolumn{3}{c|}{} & 0 & 0 & 0 \\ 
  \multicolumn{9}{c|}{\omega_{fs}k_{ps}(I_s+E_s Y_s)k_e}&\multicolumn{3}{c|}{-\omega_{fs}}& 0 & 0 & 0 \\ 
  \multicolumn{9}{c|}{} & \multicolumn{3}{c|}{} & 0 & 0 & 0 \\ \hline 
  0 & 0 & 0 & 0 & 0 & 0 & 0 & 0 & 0 & \multicolumn{1}{|c}{0} & 0 & 0 & \multicolumn{3}{|c}{ }\\ 
  0 & 0 & 0 & 0 & 0 & 0 & 0 & 0 & 0 & \multicolumn{1}{|c}{0} & 0 & 0 & \multicolumn{3}{|c}{-k_{prs}D_g }\\ 
  0 & 0 & 0 & 0 & 0 & 0 & 0 & 0 & 0 & \multicolumn{1}{|c}{0} & 0 & 0 & \multicolumn{3}{|c}{ } 
\end{array}\right] 
\left[\begin{array}{c} \Delta\omega_1(t)\\ \Delta e_{d1}(t)\\ \Delta e_{q1}(t) \\ 
                       \Delta\omega_2(t)\\ \Delta e_{d2}(t)\\ \Delta e_{q2}(t) \\ 
                       \Delta\omega_3(t)\\ \Delta e_{d3}(t)\\ \Delta e_{q3}(t) \\
                       \Delta P_{av1}(t)\\ \Delta P_{av2}(t)\\ \Delta P_{av3}(t) \\
                       \Delta P_{ref1}(t)\\ \Delta P_{ref2}(t)\\ \Delta P_{ref3}(t) \end{array} \right] \\  
+
\left[\begin{array}{ccccccccccccccc}
  0 & 0 & 0 & 0 & 0 & 0 & 0 & 0 & 0 & \multicolumn{3}{|c|}{ } & 0 & 0 & 0\\  
  0 & 0 & 0 & 0 & 0 & 0 & 0 & 0 & 0 & \multicolumn{3}{|c|}{ } & 0 & 0 & 0\\
  0 & 0 & 0 & 0 & 0 & 0 & 0 & 0 & 0 & \multicolumn{3}{|c|}{ } & 0 & 0 & 0\\
  0 & 0 & 0 & 0 & 0 & 0 & 0 & 0 & 0 & \multicolumn{3}{|c|}{B_{ds}k_{prs}A_g} & 0 & 0 & 0\\  
  0 & 0 & 0 & 0 & 0 & 0 & 0 & 0 & 0 & \multicolumn{3}{|c|}{ } & 0 & 0 & 0\\
  0 & 0 & 0 & 0 & 0 & 0 & 0 & 0 & 0 & \multicolumn{3}{|c|}{ } & 0 & 0 & 0\\
  0 & 0 & 0 & 0 & 0 & 0 & 0 & 0 & 0 & \multicolumn{3}{|c|}{ } & 0 & 0 & 0\\  
  0 & 0 & 0 & 0 & 0 & 0 & 0 & 0 & 0 & \multicolumn{3}{|c|}{ } & 0 & 0 & 0\\
  0 & 0 & 0 & 0 & 0 & 0 & 0 & 0 & 0 & \multicolumn{3}{|c|}{ } & 0 & 0 & 0\\ \hline 
  0 & 0 & 0 & 0 & 0 & 0 & 0 & 0 & 0 & \multicolumn{1}{|c}{0}& 0 &\multicolumn{1}{c|}{0} & 0 & 0 & 0 \\ 
  0 & 0 & 0 & 0 & 0 & 0 & 0 & 0 & 0 & \multicolumn{1}{|c}{0}& 0 &\multicolumn{1}{c|}{0} & 0 & 0 & 0 \\ 
0 & 0 & 0 & 0 & 0 & 0 & 0 & 0 & 0 & \multicolumn{1}{|c}{0}& 0 &\multicolumn{1}{c|}{0} & 0 & 0 & 0 \\ \hline 
  0 & 0 & 0 & 0 & 0 & 0 & 0 & 0 & 0 & \multicolumn{3}{|c|}{ } & 0 & 0 & 0\\  
  0 & 0 & 0 & 0 & 0 & 0 & 0 & 0 & 0 & \multicolumn{3}{|c|}{k_{prs}A_g }& 0 & 0 & 0\\
  0 & 0 & 0 & 0 & 0 & 0 & 0 & 0 & 0 & \multicolumn{3}{|c|}{ } & 0 & 0 & 0
\end{array}\right] 
\left[\begin{array}{c} 
\Delta\omega_1(t-t_d)\\ \Delta e_{d1}(t-t_d)\\ \Delta e_{q1}(t-t_d) \\ 
\Delta\omega_2(t-t_d)\\ \Delta e_{d2}(t-t_d)\\ \Delta e_{q2}(t-t_d) \\ 
\Delta\omega_3(t-t_d)\\ \Delta e_{d3}(t-t_d)\\ \Delta e_{q3}(t-t_d) \\
\Delta P_{av1}(t-t_d)\\ \Delta P_{av2}(t-t_d)\\ \Delta P_{av3}(t-t_d) \\
\Delta P_{ref1}(t-t_d)\\ \Delta P_{ref2}(t-t_d)\\ \Delta P_{ref3}(t-t_d) \end{array} \right]
\label{sysstate_eq}
\end{eqnarray}

\hrulefill
\vspace*{4pt}
\end{figure*}
\normalsize


The small-signal model represented by (\ref{sysstate_eq}) can be expressed representatively as (\ref{symsysstate_eq}), where $\phi(t)$ is the initial history function. Equation (\ref{symsysstate_eq}) belongs to the class of DDE with a single delay \cite{thesis_elias}.

\begin{equation}
 \left\{\begin{array}{l l}
 \Delta\dot{X}(t) = \mathbf{A} \Delta X(t) + \mathbf{A_d} \Delta X(t-t_d);  &t>0\\
 \Delta X(t)= \phi (t); &t \in [-t_d,0] \end{array} \right. 
\label{symsysstate_eq}
\end{equation}

The characteristic equation for the system described in (\ref{symsysstate_eq}) is:

\begin{equation}
  det \big(-s\mathbf{I} + \mathbf{A} + \mathbf{A_d}e^{-st_d}\big)=0
\label{charac_eq}
\end{equation}

Equation (\ref{charac_eq}) has infinite solutions, which implies that the systems represented by (\ref{symsysstate_eq}) have an infinite number of eigenvalues \cite{Elias_Tobias}.  Different  approaches have been proposed to handle DDE's, considering analytical solutions \cite{Ulsoy} or numerical solutions \cite{Bellen}.

The spectrum of scalar single delay DDE's can be determined using the Lambert $W$ function \cite{Ulsoy}. The results from the scalar case can be extended to the non-scalar cases when the matrices $A$ and $A_d$ are simultaneously triangularizable, otherwise, the solution based on the Lambert $W$ function is not applicable to the arbitrary DDE \cite{thesis_elias}. Unfortunately, the matrices $A$ and $A_d$ of the system expressed by (\ref{sysstate_eq}) are not simultaneously triangularizable.

In the present work, in order to analyze the spectrum of the single delay DDE expressed by (\ref{sysstate_eq}), a numerical approach encountered in \cite{Breda2006} is used, with the respective Matlab code as presented in \cite{thesis_elias}. The solution of the DDE is obtained by the Matlab dde23 function.

\section{Simulation and Experimental Results}
\label{sec_simula}
In order to validate the proposed small-signal model, a number of simulations and experiments were performed considering the islanded microgrid presented in Fig. \ref{Sys_3invs}, defined by the parameters shown in Table \ref{tab_exem1}. Each node is composed of a three-phase inverter with the control scheme as presented in Section \ref{sec:1}. The reader has to keep in mind that the inverter internal controllers are neglected in the proposed small-signal model. The value of transmission line impedance for the inverter 1 was considered twice the value of the impedance of the other inverters for increasing the degree of generalization.

\begin{table}[!t]
\renewcommand{\arraystretch}{1.3}
\caption{{System Parameters and Equilibrium Point}}
\label{tab_exem1}  
\centering
\begin{tabular}{|l|c|c|}
\hline
{ Variable} & { Value} & { Unit} \\
\hline\hline
Inverter LC filter - inductor        & $1.8 $  & $mH$   \\
Inverter LC filter - capacitor       & $27.0 $    & $\mu H $ \\
Load 1 = Load 2                      & $ 119 + j0 $     & $ \Omega$ \\
Line transmission - inverter 1       & $ 0.2 + j1.131$ & $ \Omega$ \\
Line transmission - inv. 2 and 3     & $  0.1 + j0.566 $ & $ \Omega$ \\
Measuring filter cut-off frequency   &                   &    \\
$(\omega_{f1}=\omega_{f2}=\omega_{f3})$ & $31.4159 $  & rad/s \\ 
Frequency droop coefficient          &                   &    \\
\ \ \       $(k_{p1}=k_{p2}=k_{p3})$ & $ 0.0004$         & rad/s/W \\
Voltage droop coefficient            &                   &    \\ 
\ \ \       $(k_{v1}=k_{v2}=k_{v3})$ & $ 0.0005$         & V/var   \\
Freq. restoration integral gain      &                   &    \\
\ \ \       $(k_{pr1}=k_{pr2}=k_{pr3})$ & $ 5  $         & W/s\\
Voltage PR controller                &          & \\
\ \ \ proportional gain $(k_{rv})$      & $0.06$    & A/V\\
\ \ \ resonant gain     $(k_{resv})$    & $40.0$      & A/V/s\\
Current PR controller                &          & \\
\ \ \ proportional gain $(k_{ri})$      & $10.0$     & V/A\\
\ \ \ resonant gain     $(k_{resi})$    & $50.0$    & V/A/s     \\
Virtual resistance  $(R_v)$             & $1.5$     & $ \Omega$    \\
Virtual inductance  $(L_v)$             & $4    $   & $mH$         \\
Apparent power                  &                    &   \\
\ \ \ inverter 1 $(P_1+jQ_1)$   & $ 442.5 - j9.7 $    & VA     \\
\ \ \ inverter 2 $(P_2+jQ_2)$   & $ 442.5 + j8.6 $    & VA   \\
\ \ \ inverter 3 $(P_2+jQ_2)$   & $ 442.5 + j8.6 $    & VA   \\
Inverter 1 output voltage $(\vec{E_1})$    & $ 230.0 \angle 0$    & V (rms), rad\\    
Inverter 2 output voltage $(\vec{E_2})$     & $ 229.99 \angle -0.0018$  & V (rms), rad\\ 
Inverter 3 output voltage $(\vec{E_3})$     & $ 229.99 \angle -0.0018$  & V (rms), rad\\ 
Nominal frequency     $(\omega)$           & $ 314.159$         & rad/s \\ 
Switching frequency                        & $ 10 $         & kHz \\ 
\hline
\end{tabular}
\end{table}

The data communication links used in the simulations are represented by the highlighted edges shown in Fig. \ref{Consensus_Sys}. The time delay in simulations were implemented using a pure delay block $e^{-t_d s}$. 

\begin{figure}[!t]
\centering
\includegraphics[width=0.9\columnwidth]{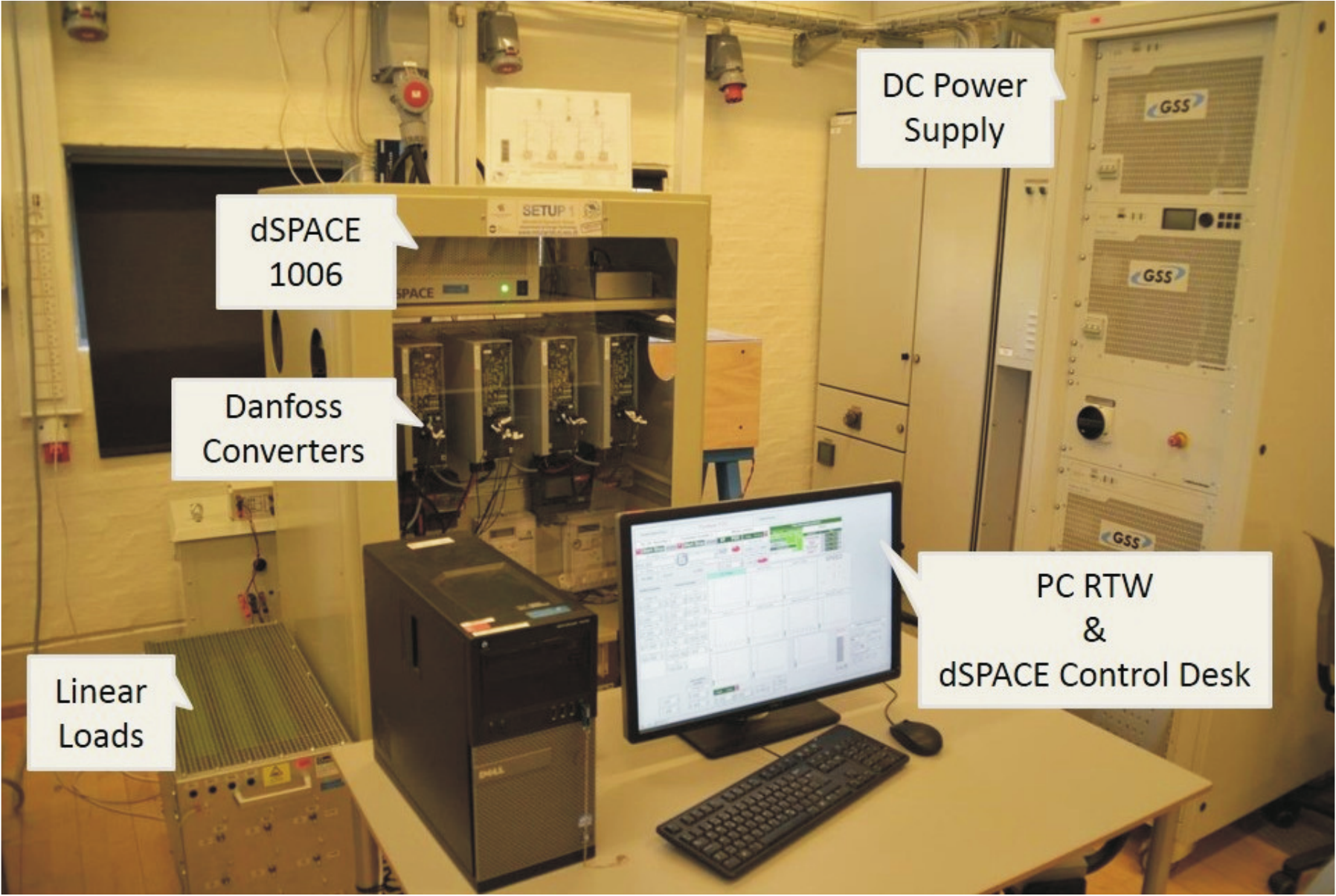}
\caption{Laboratory Setup}
\label{Lab_setup}    
\end{figure}

Each results' graph presents four curves identified as \emph{Model}, \emph{Sim1}, \emph{Sim2} and \emph{Exp} in the graph legend, which corresponds to the following results:
\begin{description}
\item[\emph{Model}] This curve corresponds to the solution of the DDE, which is a linear time-invariant system with delay in state feedback. Since the respective DDE is a small-signal model, it provides the deviations $\Delta X$, which must be added to the equilibrium point value to obtain the variable behaviour during the transient ($X=X_{eq} + \Delta X$).
\item[\emph{Sim1}] This curve is a numerical solution of the nonlinear system provided by a circuit simulator. In this case, all control blocks presented in Fig.1 are implemented, except the internal controllers, the virtual impedances and the LC filters, that is, the inverter reference voltage is equal to the inverter output voltage, and thus each inverter is an ideal voltage source.
\item[\emph{Sim2}] This curve is a numerical solution of the nonlinear system. However, in this case, the proportional-resonant controllers, the virtual impedances and the LC output filters  were included in the circuit simulator. The effect of the pulse width modulation was neglected.
\item[\emph{Exp}] This curve is an experimental result obtained from the lab prototype, as  seen in Fig. \ref{Lab_setup}. The inner loops, primary and secondary control loops were  modelled in the Matlab/Simulink and then the respective code was programmed into a dSPACE 1006 to control the three Danfoss FC302 converters. The three-unit system was powered by a Regatron GSS DC power supply. Finally, the output power and frequency of the converters were locally monitored by the dSPACE Control Desk. The inverter switching frequency was $10 kHz$.
\end{description}

In order to maintain the same comparison basis in our analysis and as the virtual impedance represents an element connected in series with the actual line impedance, both values were added to represent the inverter connection impedance to obtain the \emph{Model} and \emph{Sim1} results. This was due to the fact that the virtual impedance concept was only included in the inverter controllers for \emph{Sim2} and \emph{Exp} results.


\begin{figure*}[!hpt]
\centering
\subfloat[$\omega_1, t_d=20ms$]{\includegraphics[width=0.3\textwidth]{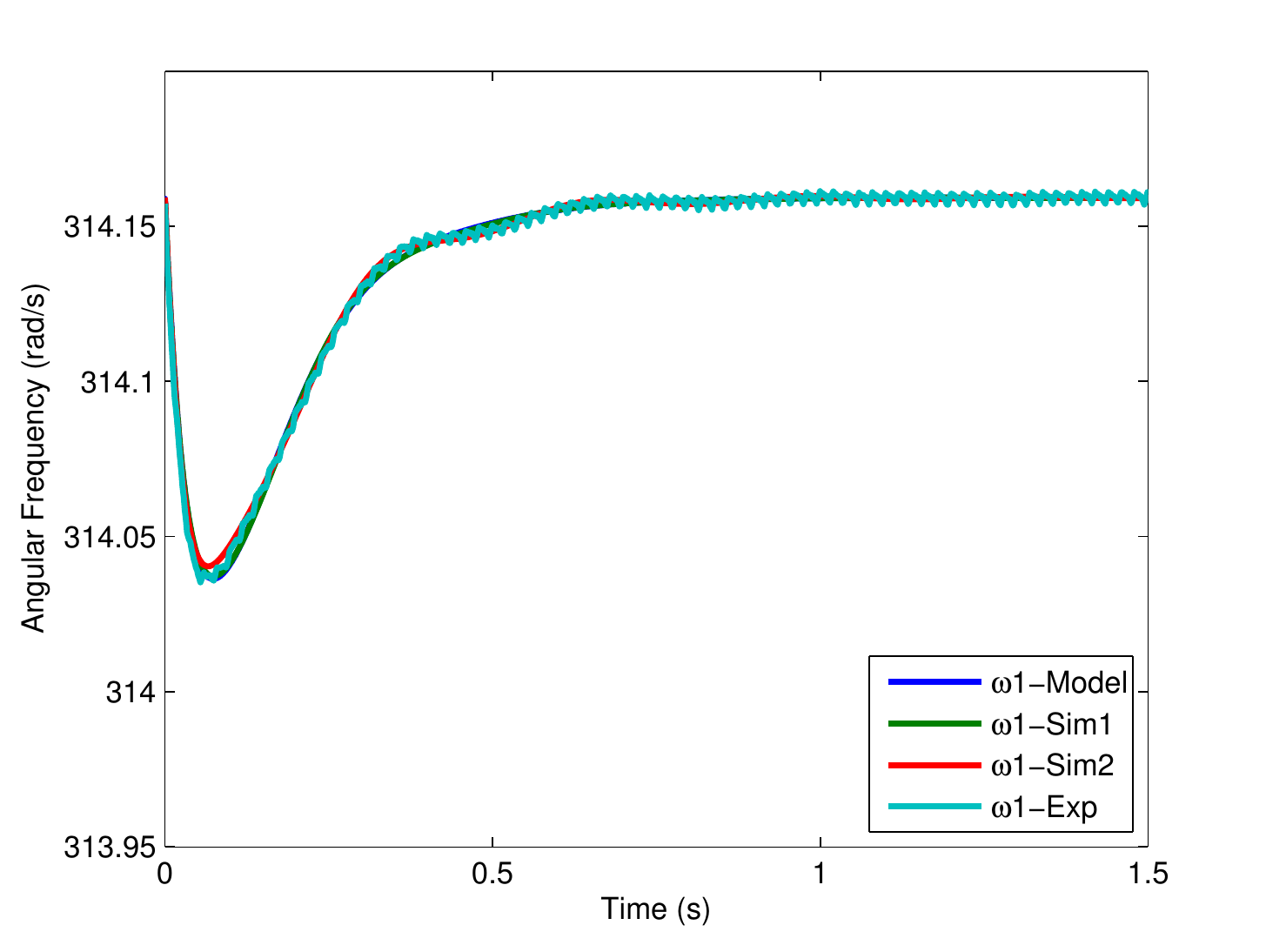}
\label{W1_20ms}}
\hfil
\subfloat[$\omega_2, t_d=20ms$]{\includegraphics[width=0.3\textwidth]{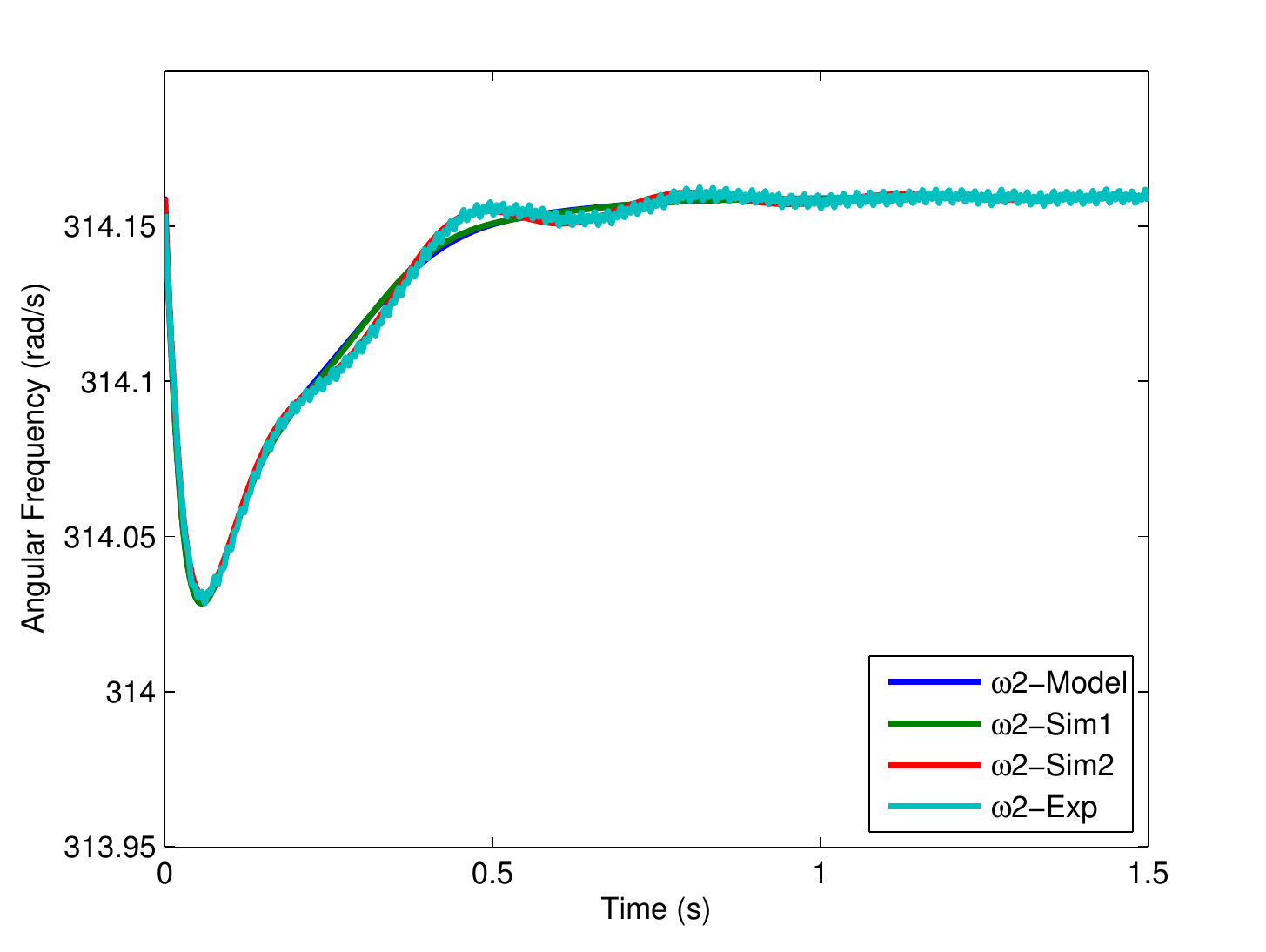}
\label{W2_20ms}}
\hfil
\subfloat[$\omega_3, t_d=20ms$]{\includegraphics[width=0.3\textwidth]{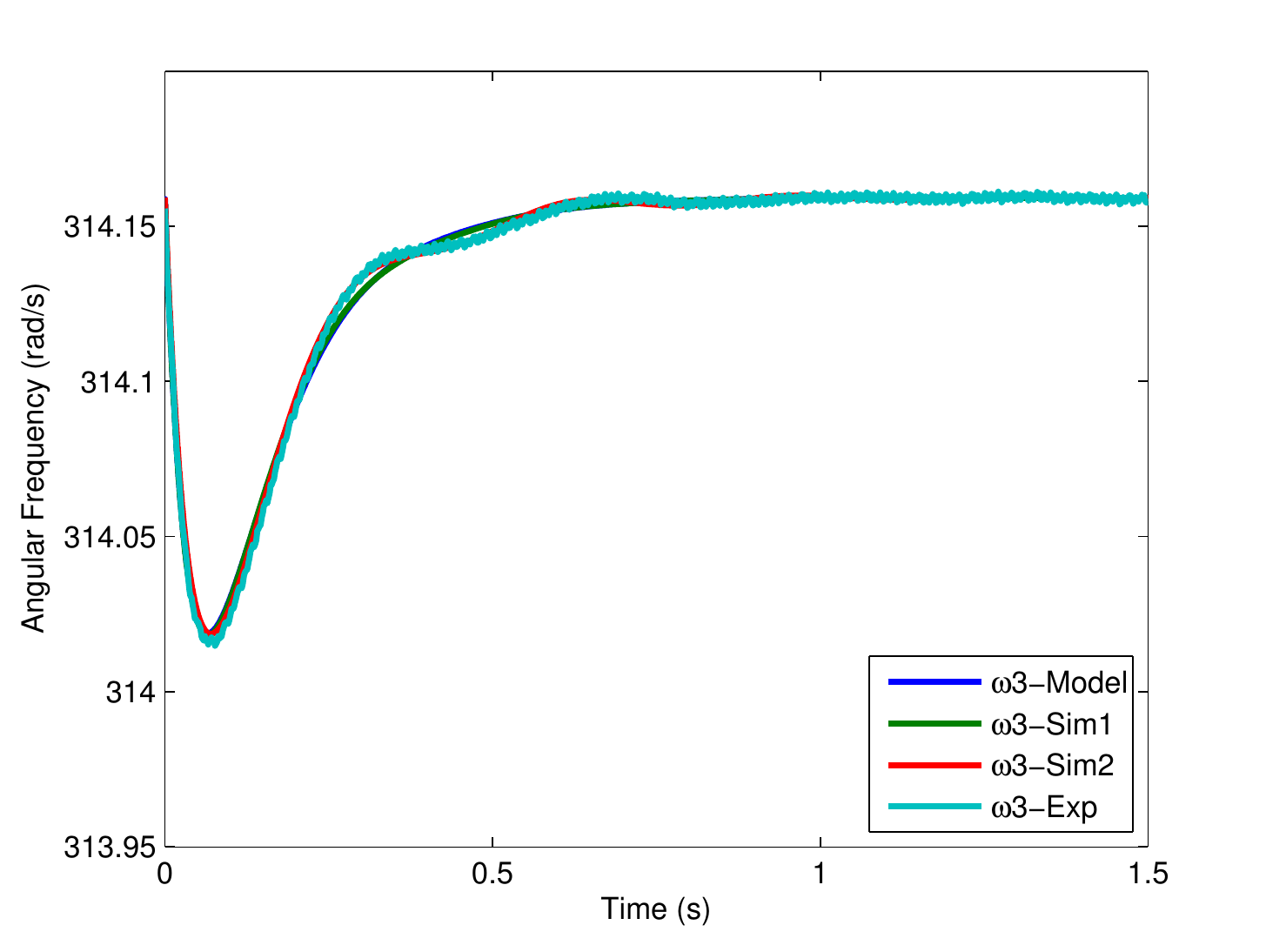}
\label{W3_20ms}}

\subfloat[$\omega_1, t_d=200ms$]{\includegraphics[width=0.3\textwidth]{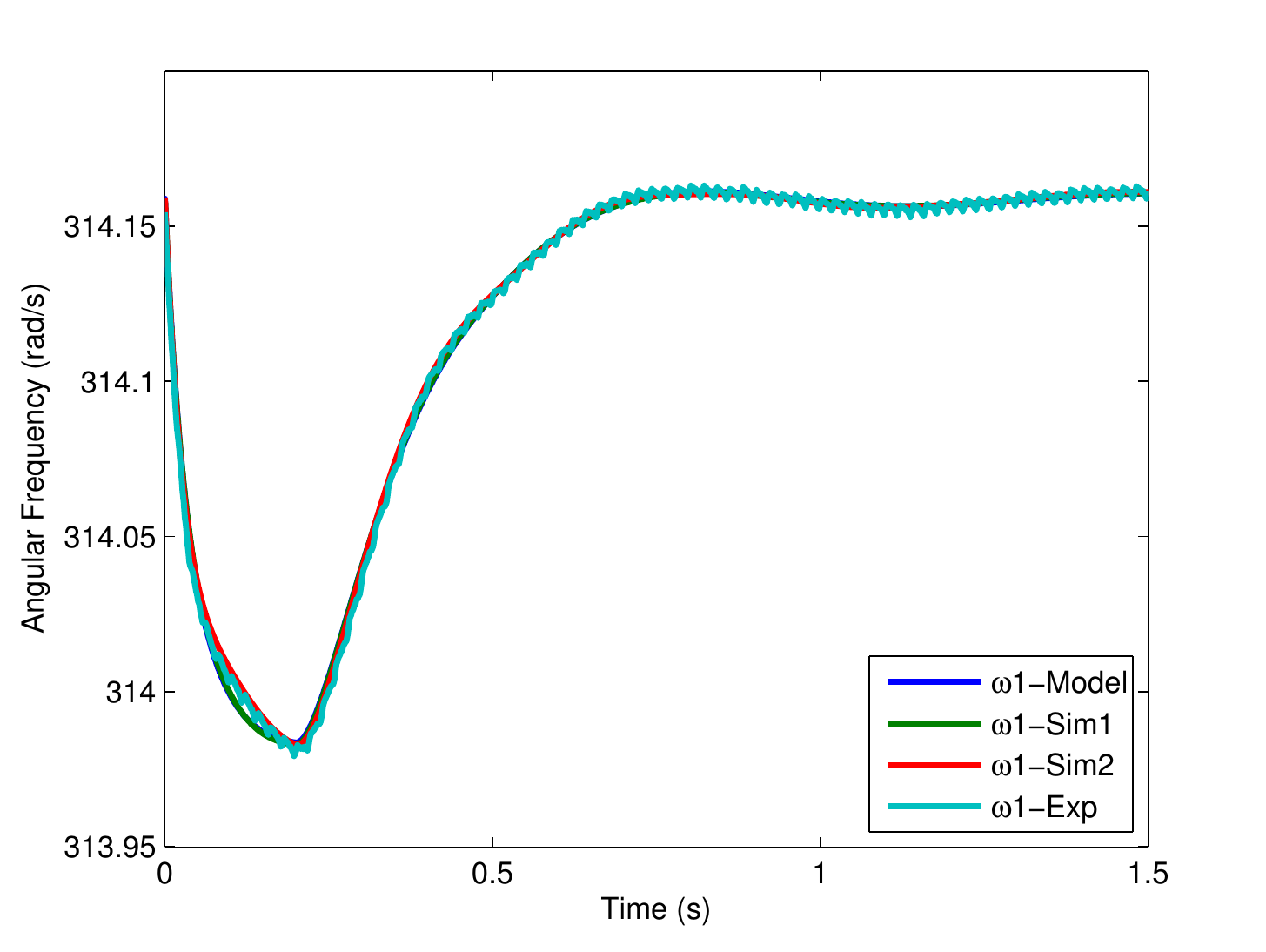}
\label{W1_200ms}}
\hfil
\subfloat[$\omega_2, t_d=200ms$]{\includegraphics[width=0.3\textwidth]{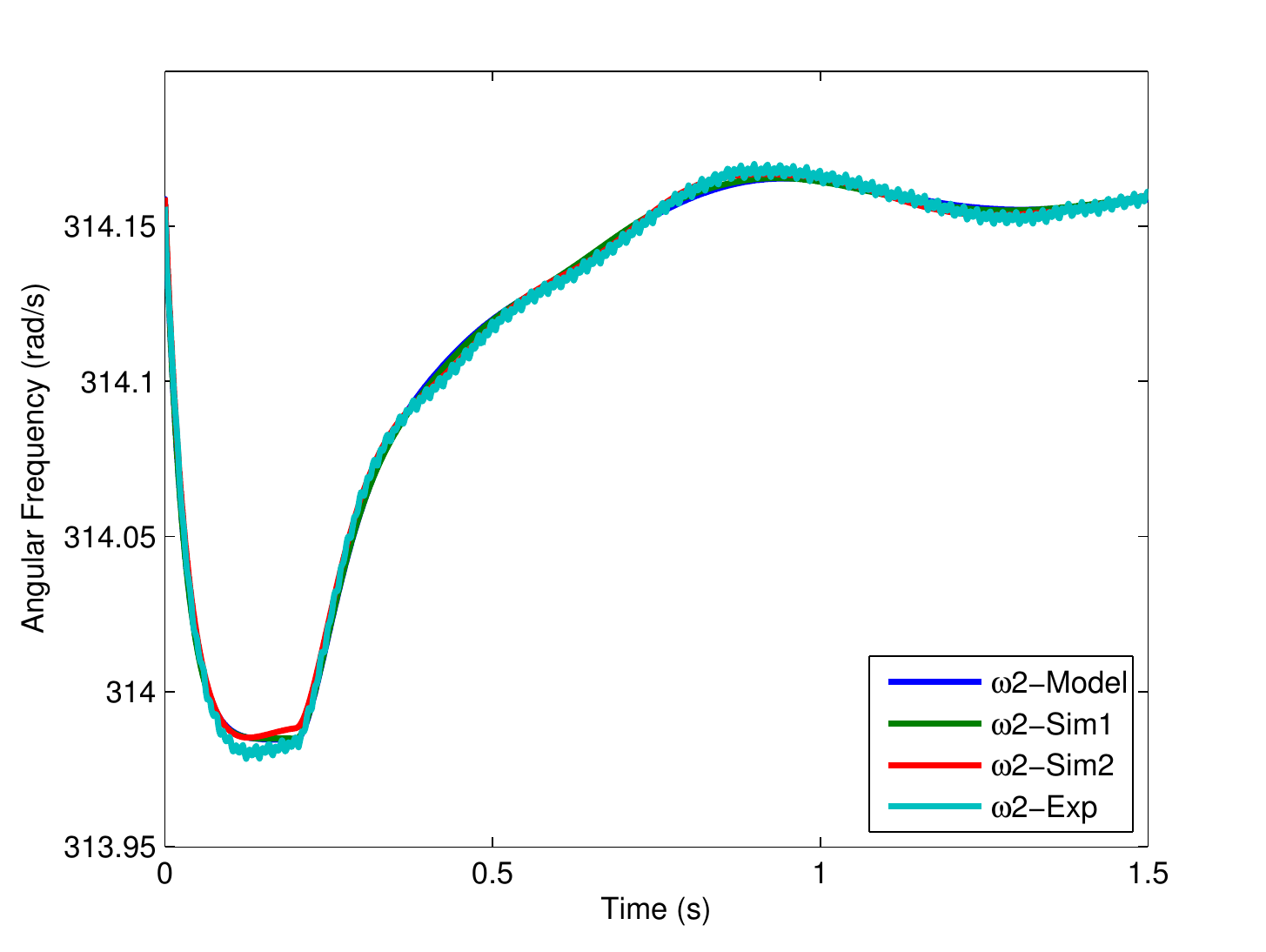}
\label{W2_200ms}}
\hfil
\subfloat[$\omega_3, t_d=200ms$]{\includegraphics[width=0.3\textwidth]{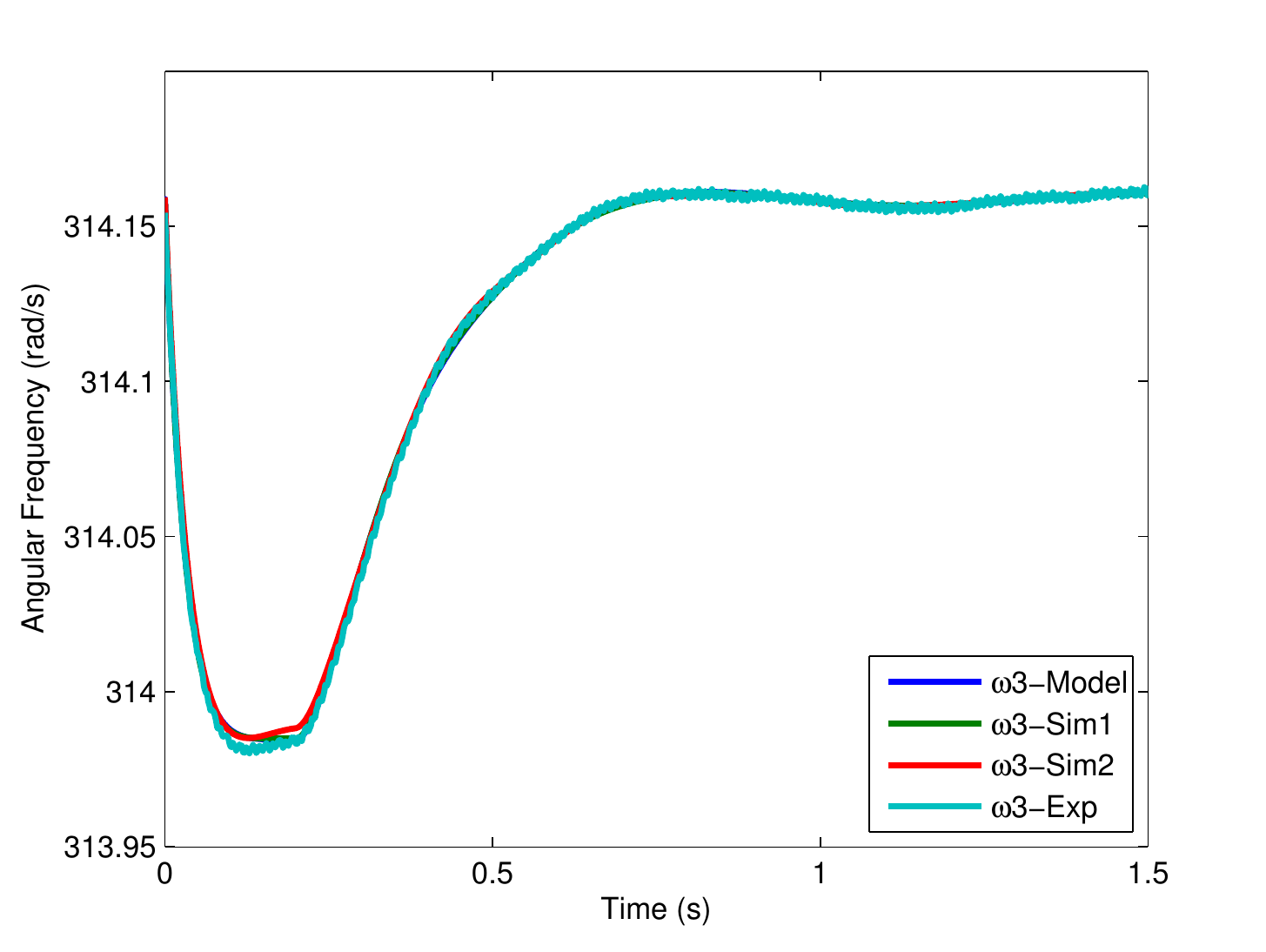}
\label{W3_200ms}}

\caption{System frequency}
\label{freq_result}
\end{figure*}


The results correspond to a transient situation between two steady states, defined by \emph{Load 1} and \emph{Load 2} (see Table \ref{tab_exem1}). Initially the system is considered as being in the steady state, as defined by the connection of \emph{Load 1}. This situation implies a constant historical function for all states ($ \Delta X(t)= \phi (t)= constant, t \in[-t_d,0] $ ) and a load flow is implemented to calculate this initial condition. Then, \emph{Load 2} is connected in parallel with \emph{Load 1} and the system moves to the new steady state, which consists of the equilibrium point shown in Table \ref{tab_exem1}. A new load flow is implemented to calculate this equilibrium point, where the parameters are used to calculate the small-signal model constants.

Fig. \ref{freq_result} shows the behavior of the frequency of the three inverters during the transient, considering two distinct values for the time delay $t_d$ in the data communication link. The frequencies were obtained by the small-signal model, by the simulations ( \emph{Sim1=ideal inverters, Sim2= real inverters}), as well as by the experiment. The  calculations for the model were obtained through the dde23 Matlab function. One notes there exists a perfect agreement between the model and simulation (\emph{Sim1}), where the inverter internal dynamics is neglected. Even considering the inverter internal dynamics, the agreement between the model, simulation (\emph{Sim2}) and the experimental result (\emph{Exp}) is very good, which shows that the inverter internal dynamics does not affect the interaction between nodes significantly and it is reasonable therefore, to neglect this interaction in the stability studies of the microgrid. 

When the load is changed, the primary control responds fast and moves the frequency of the system in order to keep the system stable and to provide load sharing. The secondary control provides the frequency restoration to the nominal value as we can see in Fig. \ref{freq_result}. At the time delay $t_d=200ms$, the system almost achieves the new equilibrium frequency, and then, even with this delay, the secondary control starts the frequency restoration.

\begin{figure}[!t]
\centering
\includegraphics[width=0.7\columnwidth]{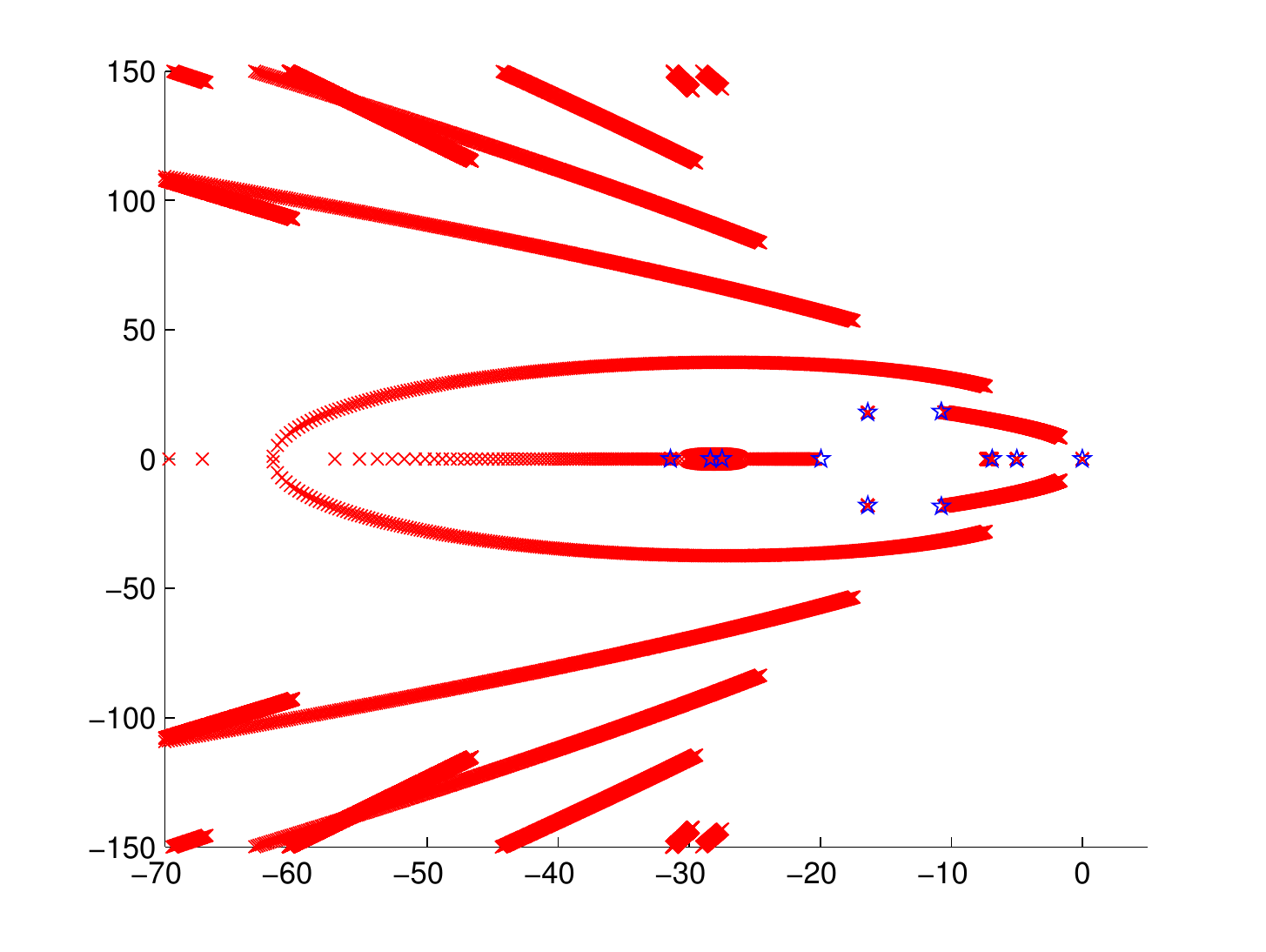}
\caption{Root locus computed with Matlab code from \cite{thesis_elias} and the number of Chebychev nodes $N=20$}
\label{Root_Locus}    
\end{figure}

The root locus plot of the system considering the time delay $t_d$ variation from 0 to 200 ms is presented in Fig. \ref{Root_Locus}, which is focused upon the rightmost eigenvalues. The finite set of eigenvalues represented by the blue stars corresponds to the system spectrum if no time delay is considered, then in this case, the system is represented by an ODE as shown by (\ref{syswithout delay}), where the $\phi(t_o)$ is the initial condition and the historical function is no longer necessary. 

\begin{equation}
 \left\{\begin{array}{l l}
 \Delta\dot{X}(t) = \big(\mathbf{A} + \mathbf{A_d} \big) \Delta X(t);  &t>0\\
 \Delta X(t_o)= \phi (t_o); &t_o = 0  \end{array} \right. 
\label{syswithout delay}
\end{equation} 

This root locus in Fig.\ref{Root_Locus} corresponds to  a numerical approximation, as it is an arduous task to determine the exact values of eigenvalues in DDE systems, mainly in the case of the presented model where $A$ and $A_d$ do not commute, that is, they are not simultaneously triangularizable. An error analysis for this numerical approach is presented in \cite{thesis_elias} for a system with an analytical solution, then it is expected that the root locus presented in Fig. \ref{Root_Locus} corresponds to a well-defined accuracy. It is noted that the system maintains stability in spite of the variation of the time delay over the considered range. As the large  time delay in communication implies a low exponential decay  in the system's answer, the low frequency modes move toward imaginary axis on the root locus graph, but they do not cross it.

\section{Extension of The Proposed Model}
\label{Ext_model}
For the sake of simplicity, a three-inverter system was considered for presenting the math developed for the proposed model and the respective validation by simulation and experimental results, as presented in Section \ref{sec:2} and \ref{sec_simula}, respectively. The proposed model can be extended in a straightforward manner to represent a microgrid with more inverters connected. For each new inverter, the model order will be increased by 5. 

In order to show an example of the model extension, in this Section, a twelve-inverter system was considered with the same droop gains presented in Table \ref{tab_exem1}. In order to increase the degree of generalization, each inverter was connected to a distinct transmission line, with inductances in the range of $0.95$ to $3.6 mH$. Across all results presented in this Section, a communication time delay $t_d$ of $200 ms$ was considered. 

\begin{figure}[!t]
\centering
\includegraphics[width=0.8\columnwidth]{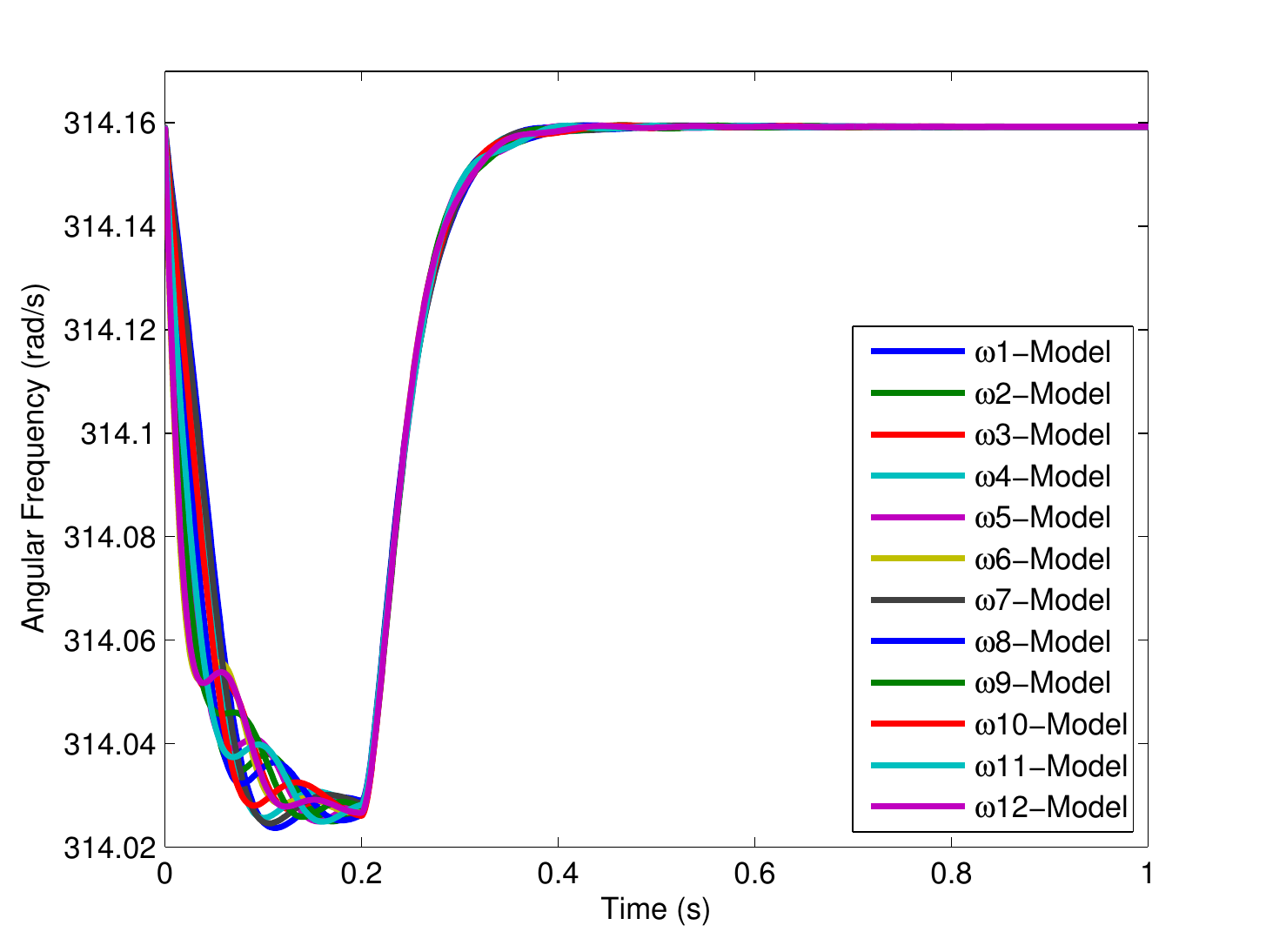}
\caption{Twelve-Inverter System Frequency - Model $t_d=200 ms$}
\label{Twelveinv_regnet_mod}    
\end{figure}

In Fig. \ref{Twelveinv_regnet_mod} the frequency of each inverter is shown during the frequency restoration process, when $Load2 (40 \Omega)$ is connect in parallel with $Load1(40 \Omega)$. This is the result of the respective $60th$ order model. In this case, a regular data communication network was used, that is, all edges in the respective twelve vertex graph are presented, which implies a fast convergence in the consensus algorithm. 

\section{ Constant Time Delay and Packet Loss in a Communication System}
\label{delay_packet}

In practice, it can be expected that a digital communication system will be used for the communication among the units. In this case, besides measurement information, packets also carry control information, which typically includes sequence numbers and/or timestamps \cite{RTP,User_Data_Protocol}. By means of buffering and inspecting sequence-numbers/timestamp information, one can insure that the receiver processes the packets received from its peers in the order that enforces equal delay on the links. This technique is commonly used in real-time communication systems, like PDH, SDH, VoIP, teleconferencing etc. Further, the buffer delay is simply incorporated in the total delay. In this sense, the delay used in the analysis in the paper could be considered as an upper limit of the total delay, made equal for all links by using standard communication techniques.

A series of experiments conducted in our lab showed that, for an off-the shelf WiFi equipment, the duration of the packet containing measurements is markedly less than $1 ms$, and the packet generation rate is of the order $1-5 ms$, which includes the transition from receiving to transmitting state. In a scenario with ca. $10$ stations, all-to-all communication and scheduled access, this implies that the frequency of secondary control can be made of the order of $50-100 Hz$.

\begin{figure}[!t]
\centering
\includegraphics[width=0.8\columnwidth]{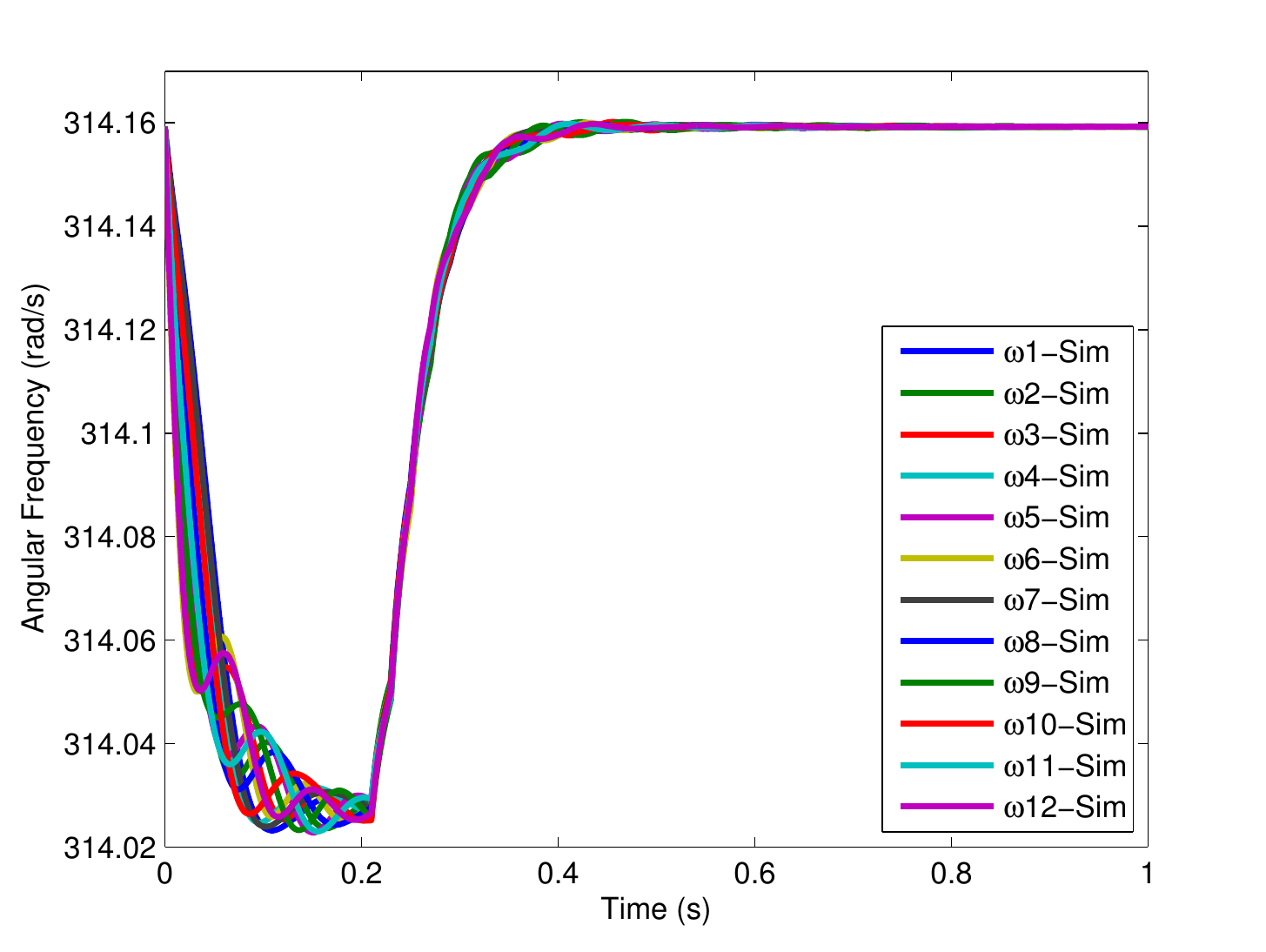}
\caption{Twelve-Inverter System Frequency - Simulation parameters: Communication Sampling Rate: 50 Hz; Packet loss probability: $10^{-2}$}
\label{Twelve_invs_w1_12_regnet_50Hz_Ploss}    
\end{figure}

In order to evaluate the performance of the secondary control considering an actual communication link, the twelve-inverter system presented in Section \ref{Ext_model} was simulated in the same transient situation. The sampling frequency of the secondary control was tuned to $50 Hz$, which is a rate that could be supported by off-the shelf equipment and considered communication setup. It was also incorporated a packet loss probability of $10^{-2}$, which can be assumed to hold for $2 Mb/s$  WiFi links in rural scenarios \cite{Barsocchi2007}. Fig. \ref{Twelve_invs_w1_12_regnet_50Hz_Ploss} shows the angular frequency of each inverter of the twelve-inverter system in the scenario described above. Compared with the result presented in Section \ref{Ext_model}, Fig. \ref{Twelveinv_regnet_mod}, one observes a good agreement. This last result shows that the usage of a realistic communication system, including the techniques mentioned above,  implies no significant difference in the system behavior. 

\section{Conclusion}
\label{conclusion}
This work presented the small-signal analysis for a microgrid system using the droop control method in the primary control and a frequency restoration function in the secondary control, where the respective communication data link was submitted to a single and constant time delay. 

The secondary control was implemented in a distributed mode, considering a consensus algorithm. The data network can be considered in different configurations, which can be easily set into the proposed small-signal model.  

The proposed small-signal model allowed for the stability analysis of a given microgrid, and it was possible to conclude that a single and constant time delay in the communication data link does not cause instability over the presented system.

In short, this work presents a  starting point for future research, since it shows a direction for dealing with time delays in the secondary control of microgrids when one considers more realistic data communication links. The assumption of a constant time delay is reasonable, even when  an actual communication system is used.  The typical sampling rate and the packet loss observed in these communication systems do not affect the performance of the secondary control in the studied microgrid.



\bibliographystyle{IEEEtranTIE}
\% Generated by IEEEtran.bst, version: 1.12 (2007/01/11)

\end{document}